\documentclass[sigconf,letterpaper,9pt]{acmart}

\usepackage{libertine}
\usepackage[T1]{fontenc}
\usepackage{graphicx,graphics}
\usepackage{url}
\usepackage{multirow}
\usepackage{color}
\usepackage{subfig}
\usepackage{balance}
\usepackage{amssymb}
\usepackage{algorithm2e}

\newcommand{\todo}[1]{\textcolor{red}{TODO: #1}}

\begin{document}

\settopmatter{printacmref=false} 		% Removes citation information below abstract
\renewcommand\footnotetextcopyrightpermission[1]{} 		% removes footnote with conference information in first column

\clubpenalty = 10000
\widowpenalty = 10000

\setlength{\belowdisplayskip}{1pt} 
\setlength{\belowdisplayshortskip}{1pt}
\setlength{\abovedisplayskip}{1pt} 
\setlength{\abovedisplayshortskip}{1pt}

%\title{Towards Fair Representation in Top-K Recommendations}
\title{Equality of Voice: Towards Fair Representation in \\ Crowdsourced Top-K Recommendations}

\author{Abhijnan Chakraborty}
\affiliation{MPI for Software Systems, Germany}
\affiliation{IIT Kharagpur, India}
 
\author{Gourab K Patro}
\affiliation{IIT Kharagpur, India}

\author{Niloy Ganguly}
\affiliation{IIT Kharagpur, India}

\author{Krishna P. Gummadi}
\affiliation{MPI for Software Systems, Germany}

\author{Patrick Loiseau}
\affiliation{Univ. Grenoble Alpes, Inria, CNRS}
\affiliation{Grenoble INP, LIG \& MPI SWS}

\begin{abstract}
To help their users to discover important items at a particular time, major websites like Twitter, Yelp, TripAdvisor or NYTimes provide Top-K recommendations (%non-personalised recommendations on home or landing page 
e.g., 10 Trending Topics, Top 5 Hotels in Paris or 10 Most Viewed News Stories), which rely on crowd-sourced popularity signals to select the items. However, different sections of a crowd may have different preferences, and there is a %huge fraction of 
large {\it silent majority} who do not explicitly express their opinion. Also, the crowd often consists of %different 
actors like bots, spammers, or people running orchestrated campaigns. Recommendation algorithms %in use 
today largely do not consider such nuances, hence are vulnerable to strategic manipulation by small but hyper-active user groups. 

To {\it fairly aggregate the preferences of all users} while recommending top-K items, we borrow ideas from prior research on social choice theory, and identify a voting mechanism called Single Transferable Vote (STV) as having many of the fairness properties we desire in top-K item (s)elections. We develop an innovative mechanism to attribute preferences of silent majority which %is an addition to 
also make STV completely operational. We show the generalizability %and applicability 
of our approach by implementing it on two different real-world datasets. Through extensive experimentation and comparison with state-of-the-art techniques, we show that our proposed approach 
%to select top-k non-personalised recommendations, 
provides maximum user satisfaction, and cuts down drastically on items %which are 
disliked by most but hyper-actively promoted by a few users.
\end{abstract}

\begin{CCSXML}
<ccs2012>
<concept>
<concept_id>10002951.10003317.10003347.10003350</concept_id>
<concept_desc>Information systems~Recommender systems</concept_desc>
<concept_significance>500</concept_significance>
</concept>
<concept>
<concept_id>10003120.10003130.10003131.10011761</concept_id>
<concept_desc>Human-centered computing~Social media</concept_desc>
<concept_significance>500</concept_significance>
</concept>
</ccs2012>
\end{CCSXML}

%\vspace{-4mm}
\ccsdesc[500]{Information systems~Recommender systems}
\ccsdesc[500]{Human-centered computing~Social media}

%\vspace{-4mm}
\keywords{Top-K Recommendation; Fair Representation; Twitter Trends; Most Popular News; Fairness in Recommendation}

\maketitle

\vspace{-4mm}
\section{Introduction}
Many websites today are deploying top-K recommendations to help their users find important items. 
For instance, social media sites like 
Twitter recommend 10 `Trending Topics' to let users know about 
%the real-time events and 
breaking news stories. Review aggregators like Yelp or TripAdvisor 
show top 10 restaurants or hotels in a particular city. News websites like CNN or NYTimes 
show 10 most viewed or most shared stories. While some of these recommendations are {\it personalized}, i.e., 
tailored to a particular user, others are {\it non-personalized} and the same items are 
recommended to all users (at least in a geographical area). 

Such recommendations implicitly rely on crowd-sourced popularity signals to select the items.
%news sources, be they
%news media like {\tt nytimes.com}, or social media like
%Facebook and Twitter, or search engines like Google, 
Recently, concerns have been raised about the potential for {\it bias} in %their automated content 
such crowdsourced recommendation algorithms~\cite{baeza2016data}. 
For instance, Google's search query autocomplete feature has been criticized for favoring
certain political %candidates over others in elections
parties~\cite{google_bias}. 
%\footnote{theguardian.com/technology/2016/dec/16/google-autocomplete-rightwing-bias-algorithm-political-propaganda}, 
%while concerns about biases in Facebook's trending topic selection have led
%a fierce debate about the need for human editorial oversight of the
%recommended trends~\cite{fb_bias}. 
%\footnote{\url{gizmodo.com/former-facebook-workers-we-routinely-suppressed-conser-1775461006},
%  \url{newsroom.fb.com/news/2016/08/search-fyi-an-update-to-trending}}.
In another work~\cite{chakraborty2017makes}, we 
showed that the majority of Twitter trends are promoted by
crowds whose demographics differ significantly from Twitter's overall
user population, and certain demographic groups (e.g., middle-aged black
female) are severely under-represented in the process.

In this paper, we propose to reimagine top-K non-personalized crowdsourced recommendations
(e.g., trending topics or most viewed news articles) %or top videos, 
as the {\it outcomes of a multi-winner election}
that is periodically repeated. We show that
the observed biases in top-K recommendations %algorithm 
can be attributed to the unfairness in the electoral system. 
More specifically in Twitter, we observe that 
during any single election cycle (5 to 15 minutes), (a) only a tiny fraction (< 0.1\%) of
the overall %social media site's 
user population express candidate (topics or hashtag) preferences, i.e., 
{\it a vast majority of voters are silent}, 
(b) some people vote multiple times, i.e., there is {\it no one 
person, one vote principle}, and (c) voters choose from several
thousands of potential candidates (topics or hashtags), splitting
their votes over several moderate and reasonable topics, and thereby,
allowing extreme topics (representing highly biased view points) to
be selected. Today's %plurality-based %({\it first-past-the-post})
trending topic (s)election algorithms are vulnerable to electing
such fringe trends with as low as 0.001\% of the electorate support.

To address the unfairness in item selections, % algorithms, 
we borrow ideas from extensive prior research on social choice theory. 
%, in general and voting theory, in particular. 
We focus on electoral mechanisms that attempt to ensure two types of fairness criteria: 
%(amongst others) in voting: 
{\it proportional representation} that
requires the divisions in (the topical interests of) the electorate
to be reflected proportionally in the elected body (i.e., selected
items) and {\it anti-plurality}, where an extremist candidate (item)
highly disliked by a vast majority of voters has little chance of
getting (s)elected. We survey existing literature and identify a
voting mechanism {\it Single Transferable Vote (STV)} as having %many of 
the properties we desire in top-K item (s)elections.

However, implementing STV-based item selection poses a technical
challenge: to deter strategic manipulation, STV requires every user to provide a preference ranking
over all candidates. Requiring the website users to rank
thousands of candidate items makes the scheme impractical. We solve
this challenge by proposing to {\it automatically infer} the %topic
preference rankings for users. Fortunately, we can leverage the rich
existing literature on personalized recommendations to 
%on ranking 
rank %topics
items according to individual personal preferences of users. 
In fact, sites like Facebook and Twitter already use personal preferences to order
topics in users' newsfeeds~\cite{fb_news_feed}. 
%\footnote{newsroom.fb.com/news/category/news-feed-fyi}. 
Additionally, our %insight 
approach enables us to account for (i.e., automatically infer the ranking choices for)
the large fraction of the electorate that is otherwise silent and
inactive during any election. % cycle.

We demonstrate the practicality and effectiveness of our ideas by
conducting a comparative analysis of different mechanisms 
for top-K recommendations using real-world data from social media site Twitter 
and news media site Adressa. % social media site. 
Over the course of a month, we collected trending topics
recommended by Twitter itself, and computed in parallel the topics that
would be recommended by four different election mechanisms including
plurality voting (where the candidates with most first place votes
win) and STV.  
At a high-level, our findings demonstrate that trending
topics elected by STV are significantly less demographically biased
than those selected by both plurality-based voting schemes and Twitter
itself. At a lower-level, our analysis reveals how the improvement
in STV selected topics arise from STV's fairness criteria of
proportional representation (which selects topics such that most users
have at least one of their highly preferred topics included in the
elected set) %of topics) 
and anti-plurality (which rejects highly biased topics disliked by a majority of %site 
users). We further evaluate the mechanisms for recommending most popular Adressa news stories 
every day throughout a two-months period, and make similar observations.

In summary, we make the following contributions in this paper: 
(a) by mapping crowdsourced recommendations to multi-winner elections, 
we show how the bias in recommendation can be traced back to the unfairness in the electoral process; 
(b) we establish the fairness properties desired in crowdsourced recommendations, and identify an electoral method, 
STV, which ensures fair representation in such contexts; 
(c) we implement STV by devising a mechanism to provide equality of voice even to the users who are otherwise silent  
during the election cycle. To the best of our knowledge, ours is the first attempt to introduce fairness in crowdsourced recommendations, 
and we hope that the work will be an important addition to the growing literature on fairness, bias and transparency of 
algorithmic decision making systems.

\vspace{-2mm}
\section{Background and Motivation}
\label{sec:motivation}
As mentioned earlier, {\it non-personalized} top-K recommendations in different websites rely on crowdsourced popularity signals 
to select the contents. For example, Twitter recommends hashtags and key-phrases as trending when their popularity 
among the crowds exhibit a sudden spike~\cite{twitter_trends}. Many news websites like NYTimes ({\tt nytimes.com}) 
or BBC ({\tt bbc.com/news}) recommend stories that are most read or most shared by their audience. 
Multiple recent works have highlighted the potential for bias in such recommendations. 

\begin{figure*}[tb]
\center{
\subfloat[{\bf }]{\includegraphics[width=0.25\textwidth]{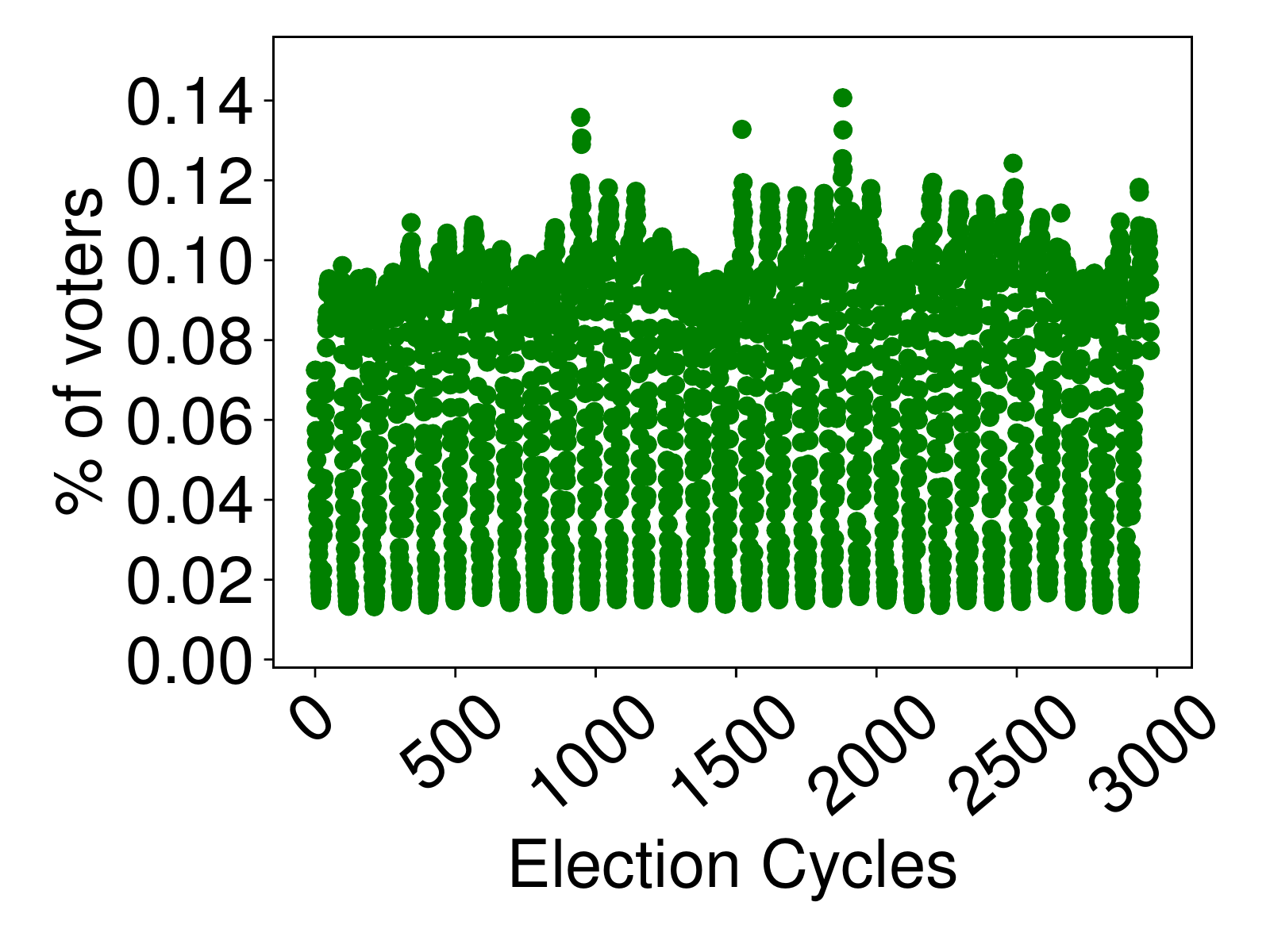}}
\hfil
\subfloat[{\bf }]{\includegraphics[width=0.25\textwidth]{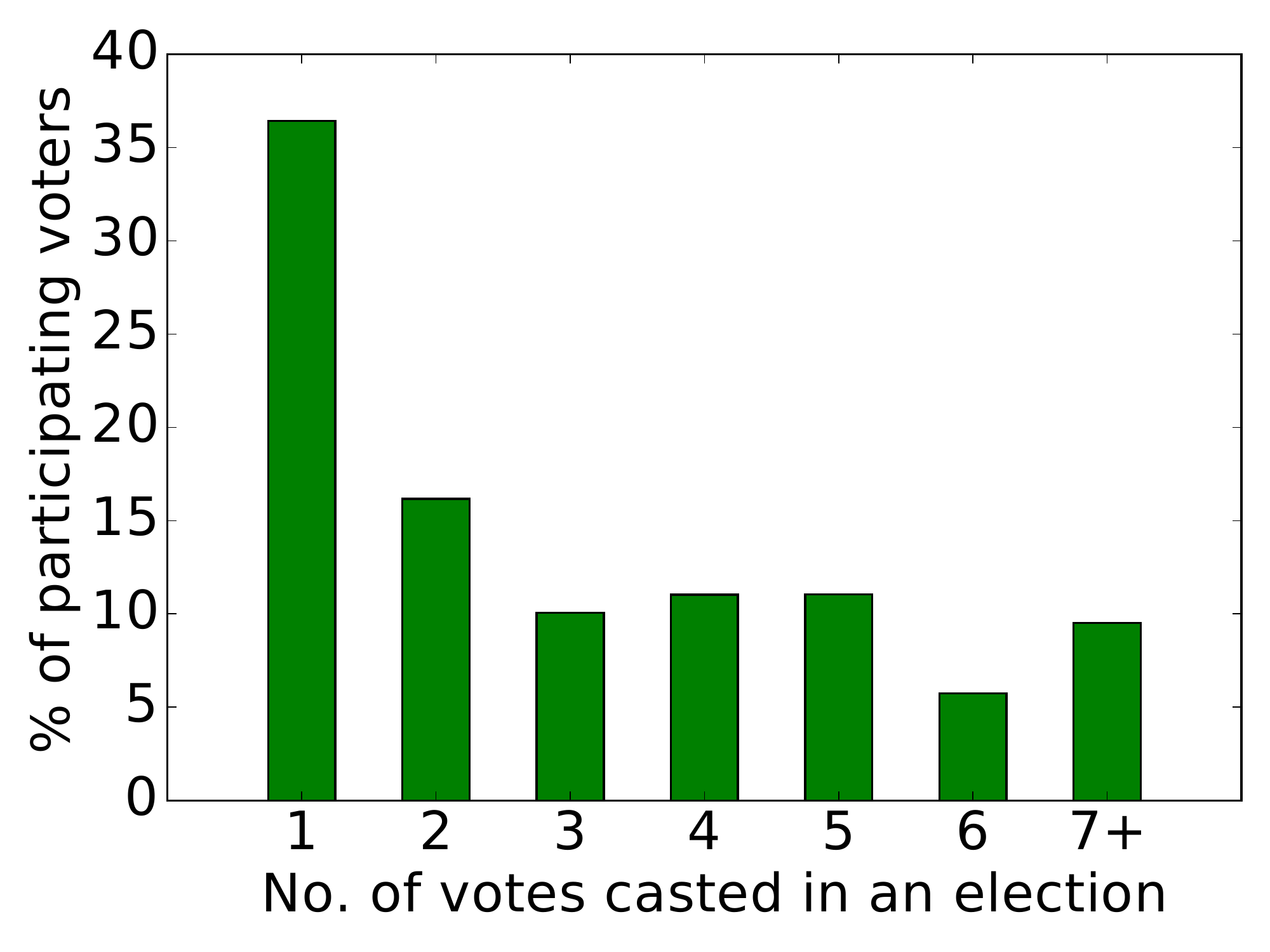}}
\hfil
\subfloat[{\bf }]{\includegraphics[width=0.25\textwidth]{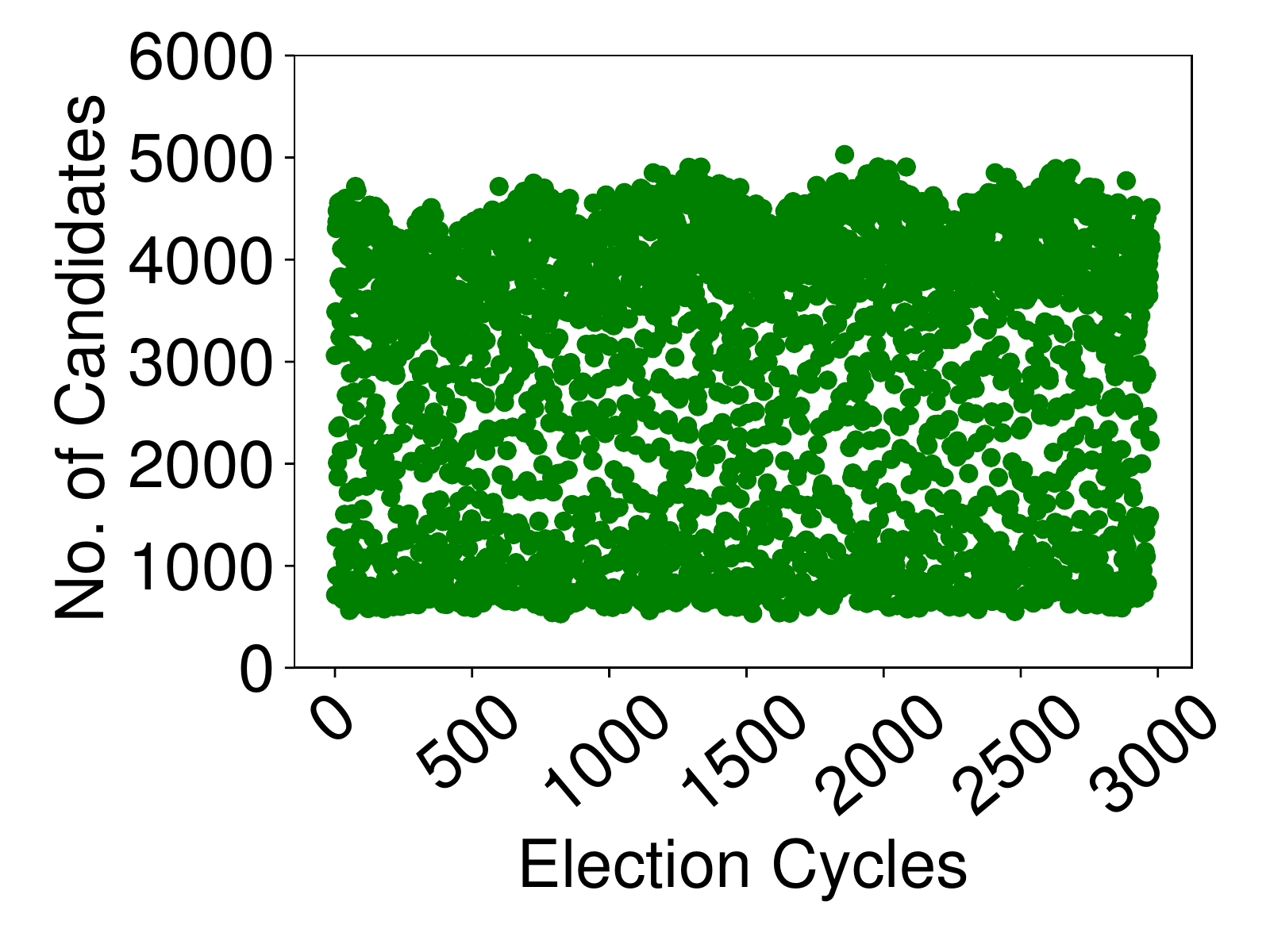}}
\hfil
\subfloat[{\bf }]{\includegraphics[width=0.25\textwidth]{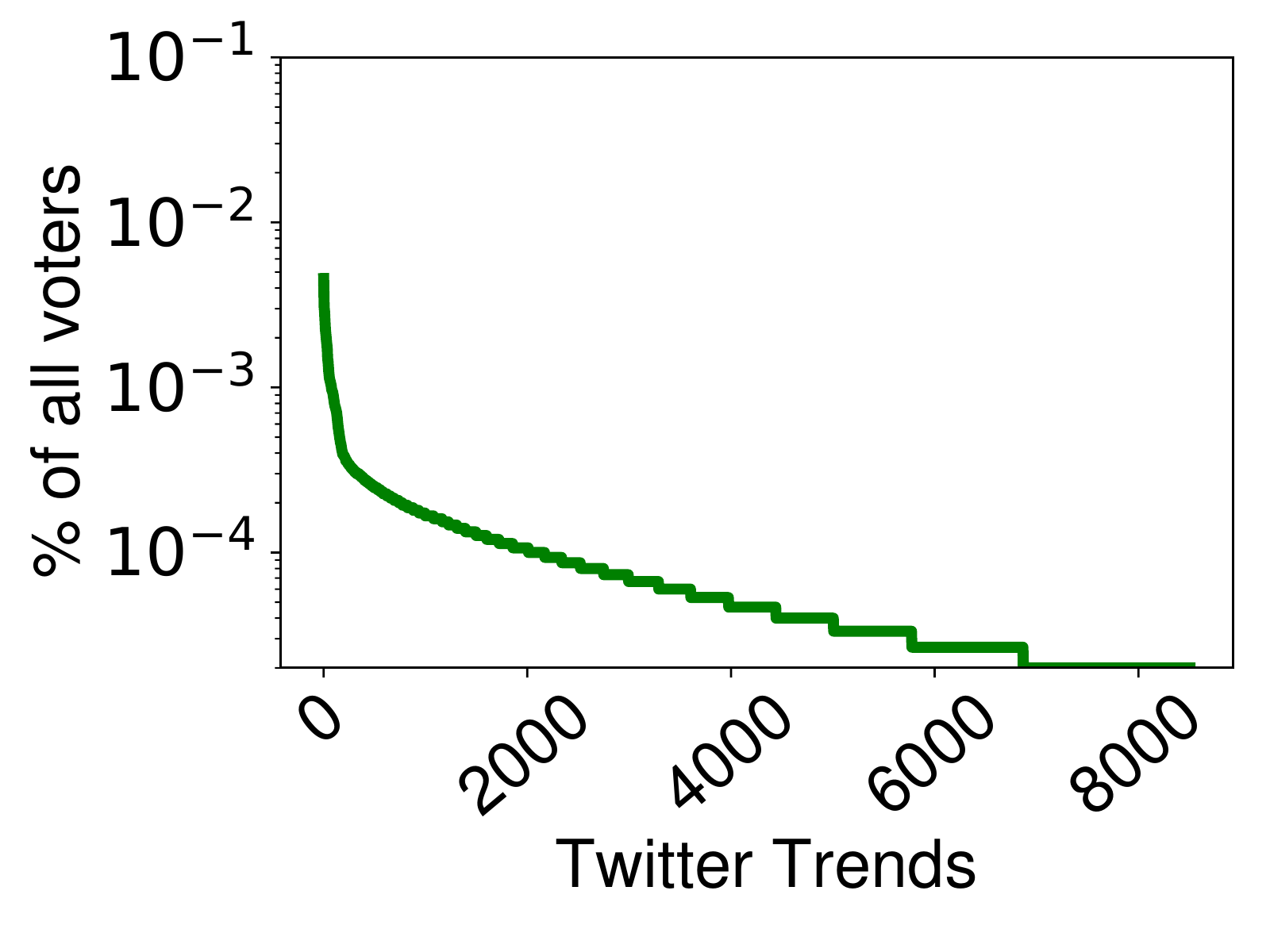}}
}
\vspace*{-4mm}
\caption{{\bf (a) Percentage of all voters participating during different election cycles in Twitter.  
(b) Average number of votes casted by different voters during an election. 
(c) Number of potential candidates for becoming trending during election cycles. 
(d) Percentage of overall population needed to make different topics trending.}}
\label{fig:active_user_stat}
\vspace*{-4mm}
\end{figure*}

\vspace{-2mm}
\subsection{Biases in crowdsourced recommendations}
%Recently, there has been a lot of debate and concerns regarding the bias in algorithms operating over big crowd-sourced data. For instance, 
Google's search query autocomplete feature has been criticized as favoring
certain political %candidates over others in elections
parties~\cite{google_bias}, 
%\footnote{theguardian.com/technology/2016/dec/16/google-autocomplete-rightwing-bias-algorithm-political-propaganda}, 
while concerns about political biases in Facebook's trending topic selection have led a fierce debate about the need for human editorial 
oversight of the recommended trends~\cite{fb_bias}. Interestingly, after Facebook removed the human editors who used to 
oversee the topics (popular among the crowds) before they were recommended to the users~\cite{fb_human_removal}, 
%\footnote{\url{newsroom.fb.com/news/2016/08/search-fyi-an-update-to-trending}}, 
it was accused of featuring fake news as trending~\cite{fb_fake_news}. 
In our earlier work~\cite{chakraborty2017makes}, we showed that the demographics of promoters of 
Twitter trends differ significantly from Twitter's overall
user population, and certain demographic groups are under-represented in the process. 
Similarly, Baker {\it et al.}~\cite{baker2013white} 
found that the gender and racial stereotypes get perpetuated in Google search auto complete suggestions.

Going beyond demographic bias, different types of actors (such as spammers,
trend hijackers or automated bots) disguise under the umbrella term `crowd'. 
As crowdsourced algorithms are driven by data generated by them, their outputs will reflect 
the biases in the composition of the crowds. A recent investigation by Politico~\cite{release_the_memo} 
%\footnote{\url{politico.com/magazine/story/2018/02/04/trump-twitter-russians-release-the-memo-216935}}
revealed that Twitter bots were largely responsible for the %trending hashtag 
trend \#ReleaseTheMemo. Multiple works have also investigated the roles of spammers and trend hijackers around Twitter trends~\cite{stafford2013evaluation,vandam2016detecting}.

We hypothesize that one of the main reasons behind the bias in
crowdsourced recommendations %such as Trending Topics is due to 
is the {\bf lack of fair representation} of various segments among the crowd considered in the algorithms. 
Using the datasets described next, we attempt to 
specifically identify the root causes behind the bias in the recommendations.

\vspace{-2mm}
\subsection{Datasets gathered}
\label{sec:dataset}
In this work, we consider %three different recommendations from different domains: 
two different recommendations:
recommendation of (i) trending topics, 
%(ii) top rated movies, 
and (ii) most popular news stories. 

\vspace{1mm}
\noindent {\bf (i) Trending Topics: } Social media sites like Twitter recommend a set of trending topics to 
help their users find happening events. We gathered extensive data from Twitter %covering $6$ months period from 
during February to July, 2017. Throughout this $6$ months period, 
we collected $1\%$ sample of all tweets posted in the US by applying the appropriate location
filters
%\footnote{developer.twitter.com/en/docs/tutorials/filtering-tweets-by-location} 
in the Twitter Streaming API~\cite{twitter_location_filter}. 
%\footnote{developer.twitter.com/en/docs/tutorials/consuming-streaming-data}.
In total, we collected $230M+$ tweets posted by around $15$ million US-based 
Twitter users throughout this %$6$ month 
period. Simultaneously, by querying the Twitter REST API~\cite{twitter_rest_api} 
%\footnote{\url{dev.twitter.com/rest/reference/get/trends/place}} 
every $15$-minutes during the month of July 2017, we collected all topics which became trending in the US. 
During this month, $10,877$ topics became trending, out of which
$4,367$ were {\it hashtags} and the rest were multi-word phrases.
For simplicity, we restrict our focus on trending hashtags in this paper. 
%and leave the analysis of trending phrases as future work.

%\vspace{1mm}
%\noindent {\bf (ii) Movie recommendation: } Movie websites like IMDB recommend a set of `Top Rated Movies'\footnote{\url{https://www.imdb.com/chart/top?ref_=nv_mv_250}} that are computed based on the ratings provided by their users. 
%In this paper, we recreate such recommendations using the user-movie rating data shared by MovieLens\footnote{\url{https://movielens.org}}, a movie recommendation platform. Specifically, we use the `MovieLens 20M Dataset'~\cite{harper2016movielens} which consists of $20M+$ ratings of $27,278$ movies given by $138,493$ users between January, 1995 and March, 2015. Using this dataset, we then simulate the recommendation of Top 25 movies at the end of every year.

\vspace{1mm}
\noindent {\bf (ii) Most Popular News: } All major news websites recommend a set of stories which are most popular (e.g., most read, most shared) among the crowd. To consider such recommendations, we use the `Adressa News Dataset'~\cite{gulla2017adressa} which consists of the news reading behavior of around $3.1$ million users on the Norwegian news website Adresseavisen\footnote{\url{https://www.adressa.no}, henceforth referred as `Adressa'} during the $3$ months period from January, 2017 to March, 2017. The dataset not only provides the information about the stories read by every user (out of total $48,486$ news stories), but also includes how much time a reader spent in each story. %Overall, the dataset contains information about $27M+$ news reading events. 
Using these reading times as popularity signal, we simulate the recommendation of $10$ most read news stories every day.

\vspace{-2mm}
\subsection{Reimagining Top-K recommendation as a multi-winner election}
In this paper, {\it we propose to see crowdsourced recommendation as the result of an election}, 
where the users vote for an item (e.g., a hashtag, a news story) by tweeting, reading,  or sharing it. 
We can think of every $x$ time interval as an election cycle (where $x$ can be any duration: $5$ minutes, 
$1$ hour, or $1$ day), the topics or stories tweeted (or read) in an interval as the candidates, 
and user activities during this interval serving as the ballots. 
The recommendation algorithm can then be viewed as an election method 
which considers these ballots and selects the winner items for recommendation. 
If only one item is to be selected (i.e., $K=1$), then it is a single winner election.
For $K > 1$, the corresponding election is multi-winner.

\subsection{Unfair election biases selection of items}
\label{sec:root_cause}
The mapping between top-K recommendation and multi-winner election allows us to
hypothesize that the bias in the recommended items originates from a lack of 
\emph{fair representation} in the underlying election mechanism. More specifically, we identify a number of potential root
causes, as discussed next. %Note that here $x$ is assumed to be $15$ minutes.

\vspace{1.5mm}
%\subsubsection{\bf Very few people are active during any interval}
\subsubsection{\bf Not everyone votes in every election cycle}
~\\ \noindent
Out of the thousands (or millions) of visitors to many websites, only a small fraction of them actively 
participate during any particular election cycle. 
For example, Figure~\ref{fig:active_user_stat}(a) shows the percentage of %fraction of %all 
Twitter users in our dataset who participated in the 
 trending topic selection during different %election 
 cycles throughout July, 2017. 
Although there are around 15 Million Twitter users in %the US 
our dataset (all of whom are eligible voters), 
we can observe from Figure~\ref{fig:active_user_stat}(a) that on average, only $0.052\%$ of them influence the trending topic selection.
Similarly, on average, %$9.28\%$ of users are active in MovieLens during any given year, and among Adressa readers, 
only $4.54\%$ of the Adressa readers read any news on a given day. Therefore, we can conclude that there is a large majority of website users who are {\it silent} during an election. 

\vspace{1.5mm}
\subsubsection{\bf One person can cast multiple votes in an election} %single
~\\ \noindent
%Even among the voters who participate in an election, they can
Different voters participating in an election may have different activity levels. 
For example, Figure~\ref{fig:active_user_stat}(b) shows the average percentage of participating 
voters who cast different number of votes during trending topic election in Twitter. We can see that 
%We can see in Figure~\ref{fig:active_user_stat}(b) that 
only $35\%$ of the voters vote once (i.e., use a single hashtag), 
and rest of the voters either vote for different candidates (by using multiple hashtags) 
or vote for same candidate multiple times. Although, we do not know for sure 
whether Twitter's Trending Topic selection algorithm considers multiple votes from the same
person, here we highlight that it may be vulnerable to such multi-voting.
We see similar trends among Adressa readers, where there is a huge variation in individual users' reading activities.

\vspace{1.5mm}
\subsubsection{\bf Too many candidates to choose from}
~\\ \noindent
%not clear - too many means there is some cognitive overload
In different websites today, the number of potential candidates for recommendations is 
much more than a user can possibly notice. 
News websites are producing hundreds of news stories everyday, and news readers have very limited time and attention.  
The problem is more acute for social media -- the amount of information generated is a lot more, and a user will 
encounter only those coming from her neighborhood, thus triggering a natural bias.

Figure~\ref{fig:active_user_stat}(c) shows the number of
candidate hashtags in Twitter during any election cycle. 
On average, at least 3,000 candidates compete 
to become trending in an election. 
Similarly, around 2,000 stories compete to become daily most popular news in Adressa. 

\vspace{1.5mm}
\subsubsection{\bf \% of voters needed to get an item selected is too low}
~\\ \noindent
As only a small fraction of voters participate in any election and their votes can get
split across a large number of candidates, 
%As a consequence to the points discussed above, 
effectively a tiny fraction of overall user population can make an item get
(s)elected. % during any election. 
% Figure~\ref{fig:active_user_stat}(d) shows the fraction of all voters behind actual Twitter trends. 
For example, Figure~\ref{fig:active_user_stat}(d) shows that most
of the Twitter trends enjoy the support of less than $0.001\%$ of
the overall Twitter population. This makes the elections vulnerable to %electing 
biased and manipulated trends. 
\vspace{-2mm}
%\section{Fair Voting: Criteria \& Mechanisms}
%\subsection{Fairness criteria for multi-winner elections}
%\subsection{Fairness criteria for topic election}
\section{Fairness criteria for Top-K recommendations}
\label{sec:fairness_notion}
In this work, we express the %selection of items in 
process in which top-K recommendations are chosen through crowdsourcing as an election mechanism. 
Extensive prior research in social choice theory %,  on electoral systems 
have identified several fairness criteria (properties) desired from the electoral systems~%\cite{elkind2017properties,faliszewski2016multiwinner,felsenthal1992normative,skowron2016axiomatic}. 
\cite{elkind2017properties,skowron2016axiomatic}. 
However, all fairness criteria are not applicable in a given context. In Section~\ref{sec:root_cause}, we identified the potential 
unfairness in the election mechanism that leads to bias in the crowdsourced recommendations. 
In this section, we propose three fairness properties 
an election mechanism should satisfy to make the recommendations fairly representative: 
(a) Equality of Voice, 
(b) Proportional Representation, and 
(c) Anti-Plurality.

%, we formalize crowdsourced topic selection as multi-winner election, and focus on the fairness properties that are most relevant in that context. Then, we introduce an electoral mechanism {\bf Single Transferable Vote (STV)} that satisfies many of the desired properties. 

\vspace{-2mm}
\subsection{Equality of Voice}
Most of the top-K recommendations in use today (e.g., `Most Viewed Stories' in news websites like {\tt nytimes.com}) can be intuitively  categorized as a particular type of electoral systems -- `Weighted Voting'~\cite{taylor1992characterization} (also known as `Plural Voting'~\cite{miller2003js}), where a single voter can vote for any number of candidates and that too multiple times. 
The candidates getting maximum votes are selected for recommendation (e.g., the stories getting maximum views regardless of users reading multiple stories or reading the same story multiple times). We saw earlier that there is a large variety in the activity levels of different users. Thus, effectively a hyper-active user can influence the election much more than a lesser-active user.

To avoid this issue, we propose that {\it the item (s)election algorithm should treat all website users (i.e., all voters) similarly, 
where no user is more privileged or discriminated to determine the set of winners}. 
In social choice, this property is known as {\it Anonymity Criterion}~\cite{skowron2016axiomatic}, and intuitively termed as `one person one vote'. 
One way to achieve this is to require the voters to specify their preferences over a set of candidates. 
In the recommendation context, we can compute these preference rankings based on the activities of the users 
(e.g., a user may post her highly preferred topic more than a lesser preferred topic). 
In addition, to give equal voice to the {\it silent} (i.e., less active) users, we also need to {\it infer} 
their ranked choices over candidate items. 
Fortunately, we can utilize the long lines of works in personalized recommendations for 
inferring different user's {\it personalized preferences} towards different items (detailed in the next section).

Let $\sigma_i(j)$ denote the preference rank user $i$ gives to item $j$ (where $\sigma_i(j) = 1$ denotes 
that $j$ is the most preferred item to $i$). $\beta_i = \{ \sigma_i(j) \; \forall j \in C \}$ denotes the preference 
ranking (i.e., ranked ballot) of the user $i$, where $C = \{c_1,c_2,...,c_m\}$ is the set of candidate items. 
Then, the top-K recommendation can be formally expressed as the 4-tuple $(C, P, f, K)$, where $P = \{\beta_1,\beta_2,...,\beta_n\}$ 
is the preference rankings from all users, and the selection algorithm is a function $f\colon C, P \rightarrow W$ which 
selects the set of $K$ winner items $W$ for recommendation from the candidate 
set $C$ (i.e., $W \subseteq C$) using the preference rankings $P$. 

%with most of the website users not participating in an election %or some users casting several votes
%and different users having very different , an election mechanism  In our context, this fairness property requires that every user's voice (i.e., preferences) should be considered (either using the active votes or by inferring their choices) and utilized while selecting the items for recommendation.
%\todo{This is a simplistic assumption, we can still make it better}
%Formally, a voting rule $R$ is {\it anonymous} if for two sets of voters having same preference rankings $P_1 = P_2$, $R(C, P_1) = R(C, P_2)$.

%\noindent{\bf ii. Neutrality. } 
%We also need that all candidates (i.e., all hashtags) are treated equally, a property known as {\it Neutrality Criterion}~\cite{skowron2016axiomatic}. For example, no particular hashtag should be favored, censored, or treated in any specific way. Such interventions can raise controversy similar to that around Facebook trend selection~\cite{fb_bias}. The tags should be elected solely based on the voters' choices. Suppose we interchange the preference ranks for two candidate tags $j_1$ and $j_2$ for all voters (i.e., $\forall i \; \sigma'_i(j_2) = \sigma_i(j_1)$) resulting in %the list of 
%preference rankings $P'$ from $P$. A rule $R$ is {\it neutral} if for all such $j_1$ and $j_2$, winners are updated accordingly (i.e., $j_2 \in R(C, P')$ if $j_1 \in R(C,P)$).

\subsection{Proportional Representation}
Even after considering everyone's preferences, due to the presence of too many candidate items, users' choices get split across %these candidates. In social choice, this is known as the problem with 
many {\it irrelevant alternatives}~\cite{arrow1950difficulty}. Furthermore, some alternatives may be very similar to each other %candidates 
(e.g., two hashtags or news stories referring to the same %topic or 
event), %, two movies may be from the same genres), 
and there the vote splitting can be %even 
sharper. 
Consequently, items preferred by only a small number of users may end up %winning an election 
being selected despite being %hated 
disliked by the majority. % of the users.

To illustrate this point, let us consider a toy example in Table~\ref{tab:example}, %, where we consider 
 depicting a $2$-winner election with $5$ candidate items and $100$ voters who rank them %se candidates 
 according to their choices. Assume that items 1 and 2 belong to a category C1, and items 3 and 4 belong to another category C2. Further assume that there is another item 5 representing an extreme opinion. We can see that although $39\%$ of the voters are interested in each of %the categories 
 C1 and C2, due to splitting of votes (especially between items 3 and 4),  
 the extreme item 5 can win in a {\it plurality vote}\footnote{where the candidates getting most first choice votes are (s)elected.}, %the election. 
 along with item 1. This is despite item 5 being disliked by $78\%$ voters who have put it as their last choice. 
To alleviate the problems illustrated above, 
we consider %the following 
two fairness criteria:  % that the topic election mechanism should satisfy.
(i) proportionality for solid coalitions, and (ii) Anti-plurality.

%\todo{The plurality vote would still be won by Tag 1 if all voters vote, so I find the sentence "Tag 5 can win in a {\it plurality vote}" not clear enough.}

\begin{table}[t]
%\small
\center
\begin{tabular}{|l||r|r|r|r|r||c|}
\hline
~ & 1st & 2nd & 3rd & 4th & 5th & Category \\
\hline
Item 1   & 30  & 20  & 30  & 17  & 3 & C1 \\
\hline
Item 2   & 9   & 30  & 15  & 38  & 8 & C1  \\
\hline
Item 3   & 20  & 20  & 25  & 30  & 5 & C2  \\
\hline
Item 4   & 19  & 30  & 30  & 15  & 6 & C2 \\
\hline
Item 5   & 22  & 0   & 0   & 0   & 78  & Extreme \\
\hline
\end{tabular}
\vspace{1mm}
\caption{\bf How 5 candidate items are present across different ranked choices of 100 users.}
\label{tab:example}
\vspace{-10mm}
\end{table}

\vspace{1mm}
\noindent {\bf Proportionality for solid coalitions} \\
%\todo{I have not understood the first one, not clear. Moreover is the second criteria a subset of the first criterion?}
We propose that  %\emph{proportional representation} criterion imposes that 
{\it in a fair top-K recommendation, the diversity of opinions in the 
user population should be proportionally represented in the recommended items}. 
%This property is specific of multi-winners elections as it cannot be satisfied in a single-winner election. 
To formalize proportional representation, we consider the criterion of {\it proportionality for solid coalitions}~\cite{Dummett84}. 
In social choice, a solid coalition for a set of candidates $C^{\prime} \subseteq C$ is defined as a set of 
voters $V^{\prime}$ who all rank every candidate in $C^{\prime}$ higher than any candidate outside of $C^{\prime}$. 
The proportionality for solid coalitions criterion requires that if 
%satisfied if  the following holds for any set of voters $V^{\prime}$: if $V^{\prime}$ is a solid coalition for $C^{\prime}$ and 
$V^{\prime}$ has at least $q \cdot \frac{n}{K+1}$ voters, then the set of winning candidates $W$ should contain at least $q$ candidates from $C^{\prime}$ (where $q \in \mathbb{N}$, $n$ %is the number of voters 
$= |P|$ and $K = |W|$). 

In the context of crowdsourced recommendations, different groups of users may prefer a group of items (e.g., hashtags or news) more than other items. Then, the proportional representation criteria means that %in a $k$-winners election, if a set of candidates $C^{\prime}$ is supported 
if a set of items is preferred by a group of users who represent a fraction $q \cdot \frac{1}{K+1}$ of the population, then at least $q$ items from this set should be selected. The quantity $\lfloor \frac{n}{K+1} \rfloor+1$ is known as {\it Droop Quota}~\cite{droop1881methods}. 
%, and the criterion is alternately termed as {\it Droop Proportionality} criterion~\cite{woodall1997monotonicity}. 
In Table~\ref{tab:example}, the Droop Quota is $34$ and hence, to satisfy proportional representation, a recommendation algorithm should select one item each from the categories C1 and C2.

%\noindent{\bf iv. Condorcet Consistency. }
%{\it Condorcet criterion}, a classical criterion in voting theory~\cite{condorcet1785essay}, was originally proposed for single winner elections, where the criterion suggests that if there exists a {\it Condorcet candidate} $c$ that is preferred to every other candidate by a majority of voters, then $c$ should be selected as the winner. %(also known as the {\it Condorcet winner}) is the person who gets more votes than any other candidate in all one-to-one comparisons. 
%Gehrlein~\cite{gehrlein1985condorcet} extended this criterion to multi-winner elections, where he defined a {\it Condorcet Committee} $C'$ as a set of $k$ candidates such that every member $c \in C'$ is preferred by a majority of voters to all candidates outside of $C$. In reality, such a committee may not exist in all elections (e.g., there is no Condorcet committee for the example in Table~\ref{tab:example}). Even when such committee exists, finding it is NP-Complete~\cite{gehrlein1985condorcet}. Thus, we relax the requirement and suggest that an election mechanism should minimize the number of candidates tags that are not selected, yet are preferred over the selected candidates by majority of the voters.

\vspace{-2mm}
\subsection{Anti-plurality}
In social choice theory, the 
{\it Majority Loser Criterion}~\cite{woeginger2003note} was proposed to evaluate single-winner elections, which requires that if a majority of voters prefer every other candidate over a given candidate, then that candidate should not be elected. We extend this criterion to top-K recommendations (and implicitly to multi-winner elections), where the \emph{anti-plurality} property intuitively mandates that no item disliked by a majority of the users should be recommended. We can formalize this criterion by requiring that no candidate item among the bottom $x$ percentile of the ranked choices for majority of the voters should be selected, where $x$ is a parameter of the definition. For example, any recommendation algorithm satisfying anti-plurality will not select Item 5 in Table~\ref{tab:example} because it is the last choice for $78\%$ of the users.
%. Formally, %we say that anti-plurality is satisfied if, for any candidate in $R(C,V)$, the 
%if a candidate $j$ is in the bottom $x$ percentile of ranked choices for at most a fraction $y$ of the population, where $x$ and $y$ ($y<0.5$) are parameters of the definition. 

%\section{Mechanisms for Fair Representation - 2 Pages}
%\section{Equal Voice Recommendations}
%\vspace{-1mm}
%\section{Fair Topic Selection\\ with Equality of Voice}
\section{Fair Top-K Recommendation with Equality of Voice}
Several electoral mechanisms have been proposed for multi-winner elections, which include {\it Plurality Voting, k-Borda, Chamberlain-Courant, Monroe or Approval Voting}~\cite{faliszewski2017multiwinner}. Subsequent research works have investigated different fairness criteria that these mechanisms satisfy~\cite{elkind2017properties,skowron2016axiomatic}. In this paper, we consider a particular electoral mechanism, {\it Single Transferable Vote (STV)}, that satisfies two fairness criteria we described in Section~\ref{sec:fairness_notion} -- proportional representation and anti-plurality\footnote{For brevity, we skip the comparison of STV with all other electoral methods. Interested readers are referred to \cite{elkind2017properties} to see how STV compares with other methods across  different fairness properties.}, and apply it in the context of crowdsourced top-K recommendations.
%However, in Section~\ref{sec:evaluation}, we compare STV with Plurality Voting among other baselines.

\begin{algorithm}[t]
\SetAlgoLined
\SetKwInOut{Input}{Input}\SetKwInOut{Output}{Output}
\SetKwComment{Comment}{$\triangleright$\ }{}
\Input{$\;$ Candidate list $C$, Preference rankings $P, K$}
\Output{$\;$ Set of $K$ winners ($W$)}
W = $\Phi$ \Comment*[r]{Start with an empty winner set}
%$\;\;$  
$dq = \lfloor \frac{|P|}{K+1} \rfloor +1$ \Comment*[r]{Droop quota}
\While{$|W|$ < $K$}
{
  	using P, assign votes to the first choice candidates \;
  	\eIf{a candidate $j$ has votes $\geq dq$}
  	{
  		W = W $\cup$ \{$j$\} 	\Comment*[r]{Add $j$ to the winner set}
  		remove $dq$ voters from P who rank $j$ first \;
  		transfer $j$'s surplus votes to the next preference of the corresponding voters \;
	   remove $j$ from all voters' preference rankings \;
   }{
   	   eliminate a candidate $c$ with the smallest tally \;
   	   redistribute $c$'s votes to its voters' next preferences \;
  }
}
\vspace{2mm}
\caption{\bf Single Transferable Vote (STV)}
\label{algo:stv}
\vspace{-4mm}
\end{algorithm}

\vspace{-3mm}
\subsection{Single Transferable Vote (STV)}
STV considers the ranked choices of all voters, %(where the voters order different candidates according to the their preferences), 
and then executes a series of iterations, until it finds $K$ winners. Algorithm~\ref{algo:stv} presents the pseudocode of the STV procedure. 
Consider the example in Table~\ref{tab:example} with $K = 2$. 
Here, Droop Quota ($dq$) is $34$; hence there is no winner in the first iteration. Item 2 gets eliminated transferring all 9 votes to Item 1 (assuming Item 1 to be those voters' second choices). In the second iteration, Item 1 wins and transfers excess 5 votes to Item 3 or Item 4 (lets assume Item 4). In the third iteration, Item 3 gets eliminated, transferring all its votes to Item 4. Finally, in the fourth iteration, Item 4 wins resulting in $W = $ \{Item 1, Item 4\}. The worst case time complexity of STV is $O(n \cdot m \cdot (m-K))$ where there are $n$ voters and $m$ candidates. However, some performance speedup is possible over the vanilla algorithm presented in Algorithm~\ref{algo:stv}.

\if 0
A single iteration operates as follows: If there is at least one candidate with a threshold plurality score (Droop Quota) $q = \frac{m}{k+1}+1$, a candidate with the highest plurality score is added to the list of winners; then $q$ voters that rank this candidate first are removed from the election, and the selected candidate is removed from all voters' preference orders. 
If the elected candidate has more votes than the quota, the excess votes are transferred to the next preference. 
If there is no such candidate, then a candidate with the lowest plurality score is removed from the election (ties are broken uniformly at random). The plurality scores are then recomputed. In this work, we used the implementation of STV in python-vote-core package (\url{github.com/bradbeattie/python-vote-core}).
\fi

By transferring votes to the preferred candidates beyond first choice,  
%consideration of voter choices beyond only first choice, and transfer of excess first choice votes, 
STV achieves proportional representation, where every selected candidate gets about $\frac{1}{K+1}$ fraction of electorate support~\cite{Dummett84}. 
%Although STV does not theoretically guarantee Condorcet consistency or anti-plurality, evaluations in Section~\ref{sec:evaluation} show that it least violates both properties compared to other methods.
Similarly, for candidates disliked by a majority of the users, unless they are preferred by at least $\frac{1}{K+1}$ fraction of all users, 
STV will not include them in the winner list, thus satisfying anti-plurality. 
More importantly, STV has been proved to be resistant to {\it strategic voting}, where determining a preference 
that will elect a favored candidate is NP-complete~\cite{bartholdi1991single}. 
Thus, STV would make it much harder for malicious actors 
%deter the malicious efforts by for spammers, bots or coordinated campaigners 
to manipulate the selection of items for recommendation. 
For example, consider the case of trending topic selection. Essentially, it would require at least $\frac{n}{K+1}$ compromised 
or bot accounts (which is a very large number considering $n$ = number of all Twitter users and $K = 10$) to make one topic trending.

%\todo{I find the explanation for anti-plurality less rigorous. Anti-plurality depends on $x$ so what is the precise meaning of "it satisfies anti-plurality"?}
%\todo{For resistance to strategic behavior: I understand what you mean but it is "for a fixed number of bots". It is still possible, instead of voting many times to create many accounts that vote for a topic to promote it. Worth mentioning? (not sure given space)}

%In the previous section, we saw that STV if applied in Trending topic selection can enable fair topic selection. 
Although, STV satisfies the fairness properties required in crowdsourced item selection, 
it considers the ranked choice over all candidates for every user\footnote{Though it is technically possible to apply STV with incomplete preference rankings, STV guarantees strategyproofness only when the preferences are complete~\cite{bartholdi1991single}.}, 
which gives rise to the following two problems that hinder the applicability of STV in recommending items: \\
(i) A large majority of the users do not vote during an election, and 
(ii) Even for the users who participate, it is not possible to get the ranked choice over all candidate items 
(because they may vote for only a few candidates during an election). \\
Next, we propose approaches to circumvent these two issues, enabling us to {\it guarantee equality of voice to everyone} (including silent users) and apply STV for selecting items in top-K recommendations.

\vspace{-3mm}
\subsection{Getting preference rankings of all users}
Intuitively, we can think of getting the ranked choices of a user $u$ as determining how interested $u$ is in different candidate items. 
Then, the problem gets mapped to {\it inferring user interests in personalized item recommendations}, and there is a large body of works on the same, which can be categorized into two broad classes: content based methods~\cite{lops2011content} and collaborative filtering~\cite{koren2015advances}. We first attempt to get the personalized ranked choices of Adressa readers, by applying a collaborative filtering approach based on Non-negative Matrix Factorization (NMF), as described next.

\vspace{1mm}
\noindent {\bf Inferring preferences for Adressa readers} \\
As mentioned in Section~\ref{sec:dataset}, the Adressa dataset contains information about the time different readers spent on different news stories. We first convert this {\it implicit} feedback to {\it explicit} ratings by normalizing with respect to both users' reading habits and the length of different articles. 
If a user $u$ spent $v_{u,i}$ time reading news story $i$, then we compute the normalized view duration $nv_{u,i}$ as
\begin{equation}
nv_{u,i} = \frac{v_{u,i}}{\mu_i}
\end{equation}
where $\mu_i$ is the average time spent by all users reading story $i$. Note that this normalization removes the bias of having possibly longer view duration for lengthier articles.

Once $nv_{u,i}$ values are computed for different stories for the user $u$, we divide them into $5$ quantiles and convert them into ratings. For example, the top $20$ percentile $nv_{u,i}$ values are converted to rating $5$, the next $20$ percentile values to rating $4$, and so on. We apply this mapping to every user-news interaction and end up with a user-news rating matrix $R$ where $R_{u,i}$ denotes the rating of news story $i$ computed for user $u$.

Matrix factorization approaches map both users and news stories into a joint latent feature space of $Z$ dimensions such that 
the interaction between the users and the news stories are modeled as inner products in that space. For example, 
if the vectors $x_u$ and $y_i$ denote the latent feature vectors for user $u$ and news story $i$, 
then the estimated rating $\hat{r}_{u,i}$ for a given user $u$ and a story $i$ is given by the scalar product:
\begin{equation}
\hat{r}_{u,i} = x_u^T \cdot y_i
\end{equation}

The challenge then is to find the latent feature vectors by observing the existing ratings. 
Multiple approaches have been proposed to efficiently optimize the following objective~\cite{koren2015advances}
\begin{equation}
\label{eq:nmf}
min \sum_{(u,i) \in \kappa} (r_{u,i} - \hat{r}_{u,i})^2 + \lambda (|| x_u ||_2 + || y_i ||_2)
\end{equation}
where $\kappa$ is set of user-news pairs for which the ratings are known.

In this paper, we apply the Non-negative Matrix Factorization approach proposed by Luo {\it et al.}~\cite{luo2014efficient} which solves
Equation~\eqref{eq:nmf} by using stochastic gradient descent with non-negativity constraints on the feature values. Once we get the feature vectors for different users and news stories, then the ratings can be predicted even for the unread stories. Thereafter, we compute preference ranking for the users based on the predicted and actual ratings (with actual ratings getting precedence and ties being broken randomly).
 
\vspace{1mm}
\noindent {\bf Inferring preferences for Twitter users} \\
To infer Twitter users' preferences, we considered both content based recommendation and collaborative filtering: \\
(i) Compute content based similarity between a user $u$ and hashtag $h$ by considering the set of all tweets posted by $u$ and the set of tweets containing $h$. However, we found that most users do not post enough tweets, and thus we can not accurately compute the content based similarity between a user and the candidate hashtags. \\
(ii) As there is no explicit rating available, we tried to apply a collaborative filtering based approach to compute personalized ranking using implicit feedback like favoring or retweeting~\cite{rendle2009bpr}. However, two independence assumptions in such approaches -- items are independent of each other and the users act independently -- do not hold in the context of Twitter. Hashtags are often related~\cite{samanta2017lmpp}, and Twitter users often influence other users. Further, the amount of implicit feedback is very low (in our dataset, only $21\%$ tweets get any retweets, or likes or favorites) and the set of hashtags are constantly changing. Hence, the collaborative filtering approaches could not be applied in this context. 

%In order 
To circumvent these %above mentioned 
difficulties, we utilize prior works involving topical experts on Twitter~\cite{ghosh2012cognos,bhattacharya2014inferring,zafar2016wisdom}. %by Bhattacharya {\it et al.}~\cite{bhattacharya2014inferring} and Zafar {\it et al.}~\cite{zafar2016wisdom}. 
Using the `List' feature in Twitter, users can create named groups of people they follow. 
By giving meaningful names to the lists created, they implicitly describe the members of such groups. 
Ghosh {\it et al.}~\cite{ghosh2012cognos} gathered these list information from a large number of Twitter users, and 
identified thousands of topical experts on Twitter, where the topics are very fine-grained. Then, both Bhattacharya {\it et al.}~\cite{bhattacharya2014inferring} and Zafar {\it et al.}~\cite{zafar2016wisdom} utilized these topical experts to infer interest of a particular user as well as topic of a particular 
hashtag. The basic intuition of~\cite{bhattacharya2014inferring} is that if a user is following multiple experts in some area, then he is 
likely to be interested in that area. Similarly, if multiple topical experts are posting some hashtag, then the probability that the hashtag belongs to that topic is very high~\cite{zafar2016wisdom}.

Implementing the approaches proposed in~\cite{bhattacharya2014inferring} and~\cite{zafar2016wisdom}, for a user $u$, we infer an interest vector $I_u$ considering the experts $u$ follows, and similarly, we compute a topic vector $T_h$ for a hashtag $h$ by taking into account the experts tweeting $h$. Then, for every user $u$, we normalize the interest topics in $I_u$ such that every entry in $I_u$ lies between $0$ and $1$, and all entries sum to $1$. Similarly, for every hashtag $h$, we calculate the tf-idf scores over the topics in $T_h$. We repeat this process for every user and every candidate hashtag during an election. Finally, we compute the preference scores between all users and all candidate hashtags as 
\begin{equation}
A = U \times T \times H^T
\end{equation}
where $A_{n \times m}$ is the User-Hashtag Affinity Matrix with $A_{v,h}$ denoting affinity between user $u$ and hashtag $h$; 
$U_{n \times t}$ is the User-Interest Matrix with $U_{u,j}$ representing normalized interest of $u$ in some interest topic $j$; 
$T_{t \times t}$ is the Interest-Topic Similarity Matrix,
$T_{i,j}$ representing the similarity between two topics $i$ and $j$ (we compute $T_{i,j}$ as the Jaccard Similarity between the set of experts in topic $i$ and $j$ respectively). Finally, $H_{m \times t}$ is the Hashtag-Topic Matrix where $H_{h,j}$ denotes tf-idf of topic $j$ in hashtag $h$.

%Where\\
%m = Total Number of Users\\
%t = Total Number of Topics\\
%n = Total Number of Hashtags\\

%\todo{$R$ denoted a voting rule earlier in the paper. But make a notation change only if you are sure to have time to check consistency across the paper.}

Using $A$ computed above, we can get the preference ranking of any user over the candidate hashtags. If a user $u$ participates in an election and votes for tag $h$, then $h$ is considered as the top choice in $u$'s preference ranking and other ranked positions are shifted accordingly. If a user votes for $k$ hashtags, top $k$ positions are assigned to these $k$ candidates according to their usage frequency.

\vspace{1mm}
\noindent {\bf Accuracy of the preference inference} \\
For inferring the preferences of Adressa readers, we attempted another technique based on Singular Value Decomposition (SVD)~\cite{paterek2007improving}. Comparing the Root Mean Squared Error (RMSE) between the actual ratings and ratings inferred by both SVD and NMF based approaches, we found that the NMF based approach (RMSE: $0.87$) works better than the SVD based approach (RMSE: $0.97$).

In Twitter, there is no ground truth rating or ranking. Hence, to check the accuracy of the inference of Twitter users' preference rankings, we asked $10$ volunteers (who are active Twitter users) to rank $10$ hashtags during $15$ election cycles. Then, we compute their preference ranking using our approach and checked Kendall's rank correlation coefficient 
$\tau$~\cite{kendall1955rank} between the inferred and actual rankings for every volunteer. We find the average $\tau$ value to be $0.702$ (compared to $0.317$ for random ordering), which suggests that our method can infer the ranked choices of users reasonably well.

%\niloy{The issue which stays is in a topic there can be hashtags with conflicting interest - this would be wrongly voted. I am not sure whether 
%people will buy it. Anyway not much to do now perhaps. }

\begin{figure*}[t]
\center{
\subfloat[{\bf }]{\includegraphics[width=0.24\textwidth]{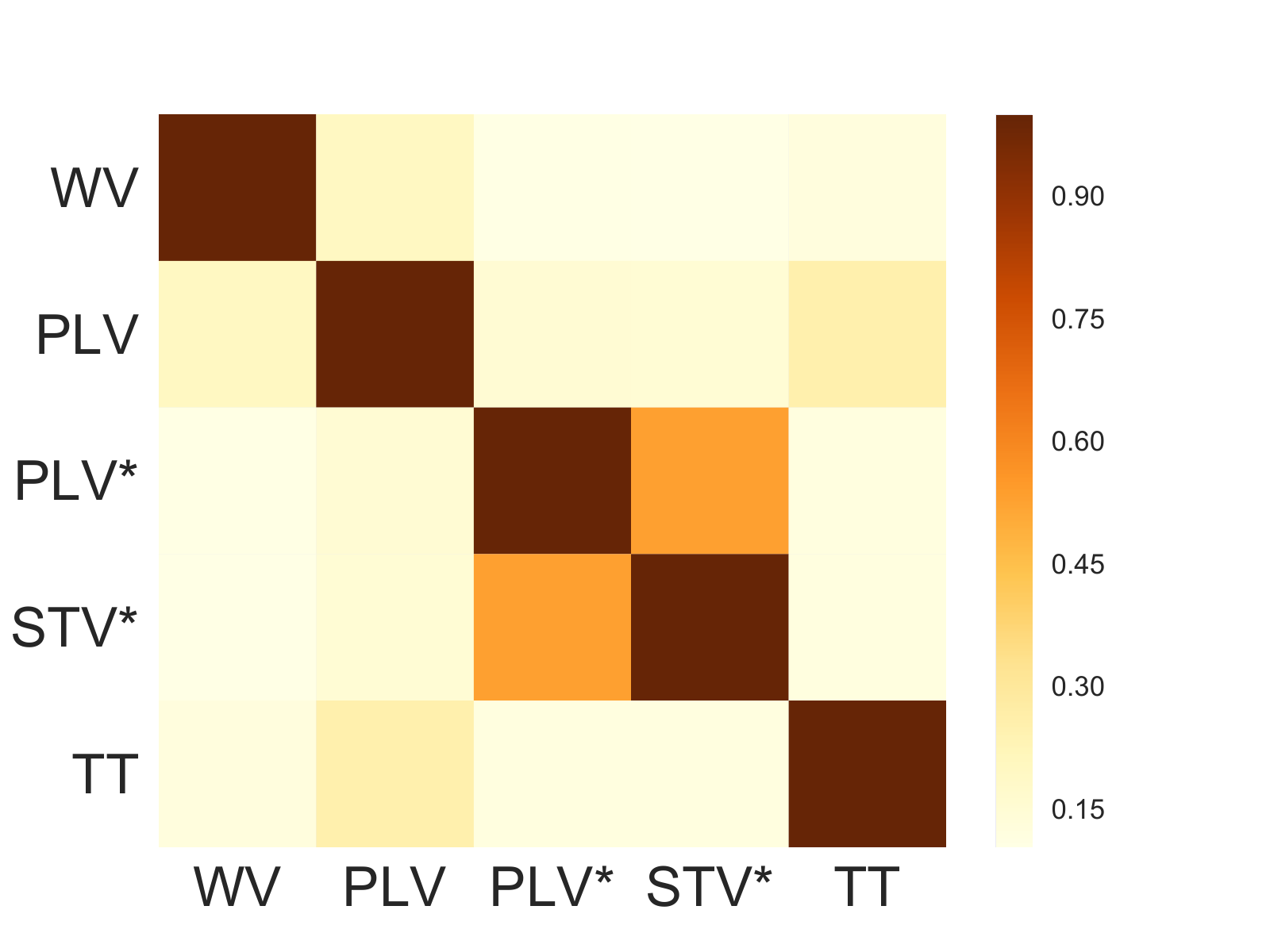}}
\hfil
\subfloat[{\bf }]{\includegraphics[width=0.24\textwidth]{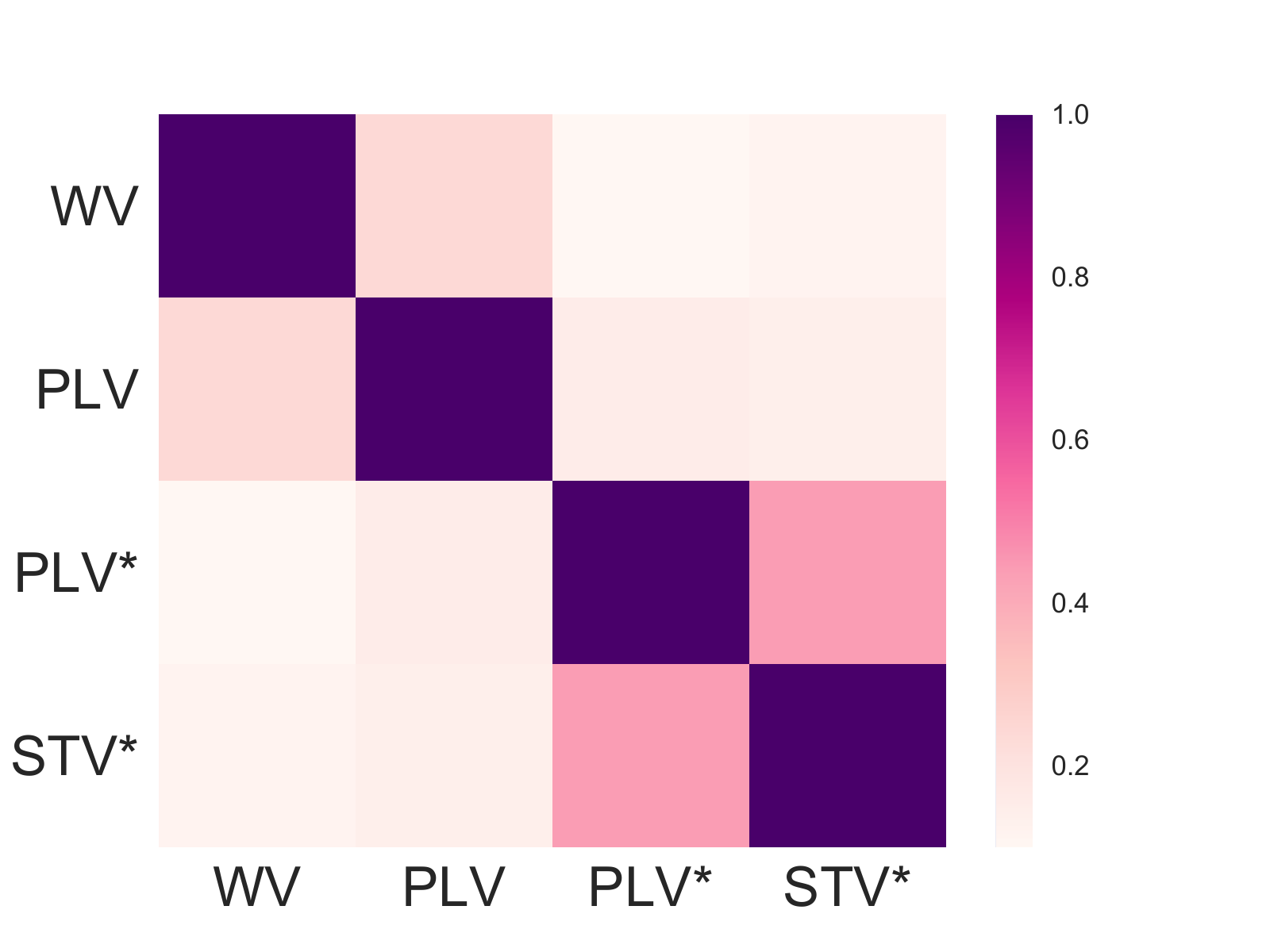}}
\hfil
\subfloat[{\bf }]{\includegraphics[width=0.24\textwidth]{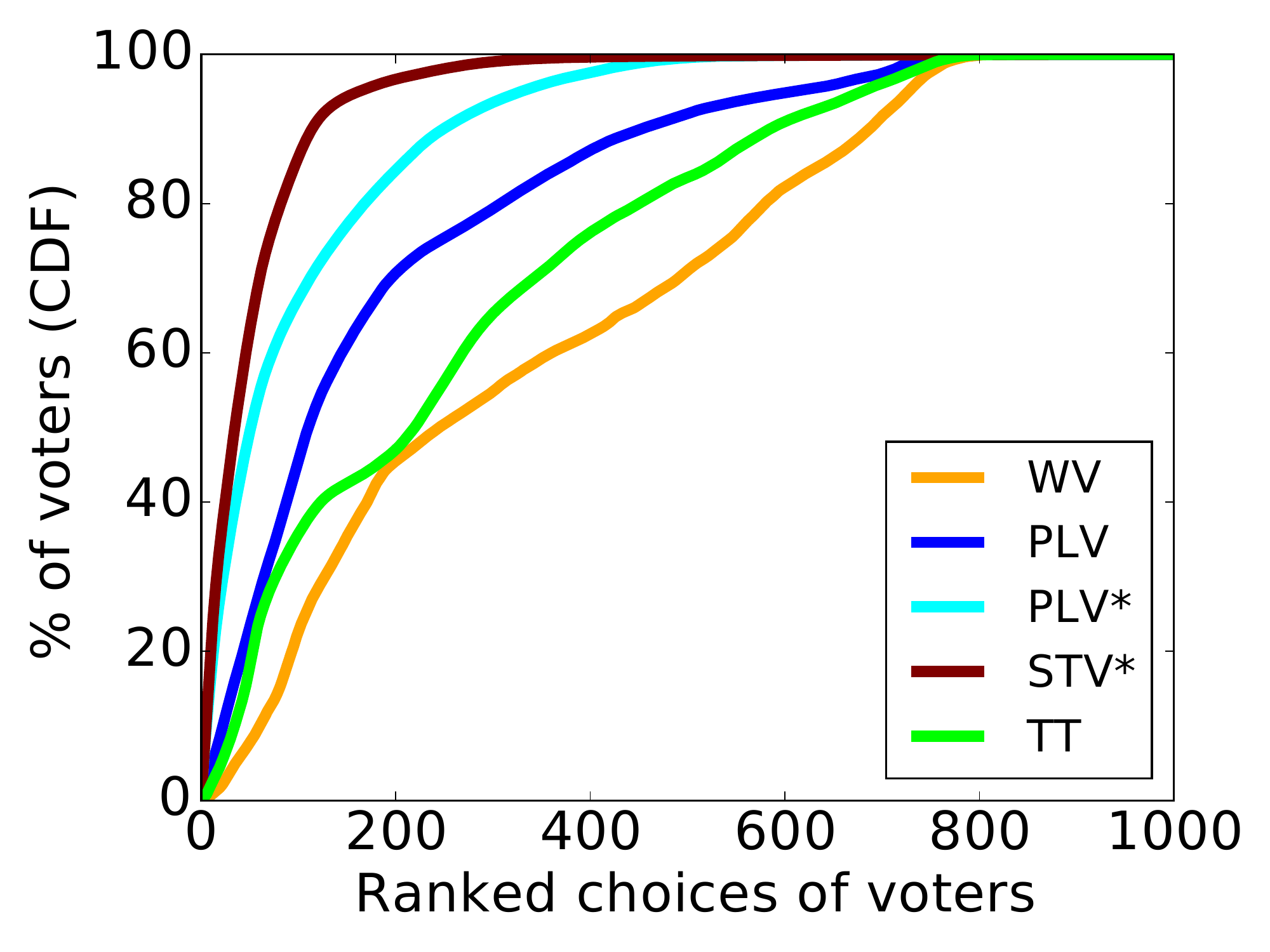}}
\hfil
\subfloat[{\bf }]{\includegraphics[width=0.24\textwidth]{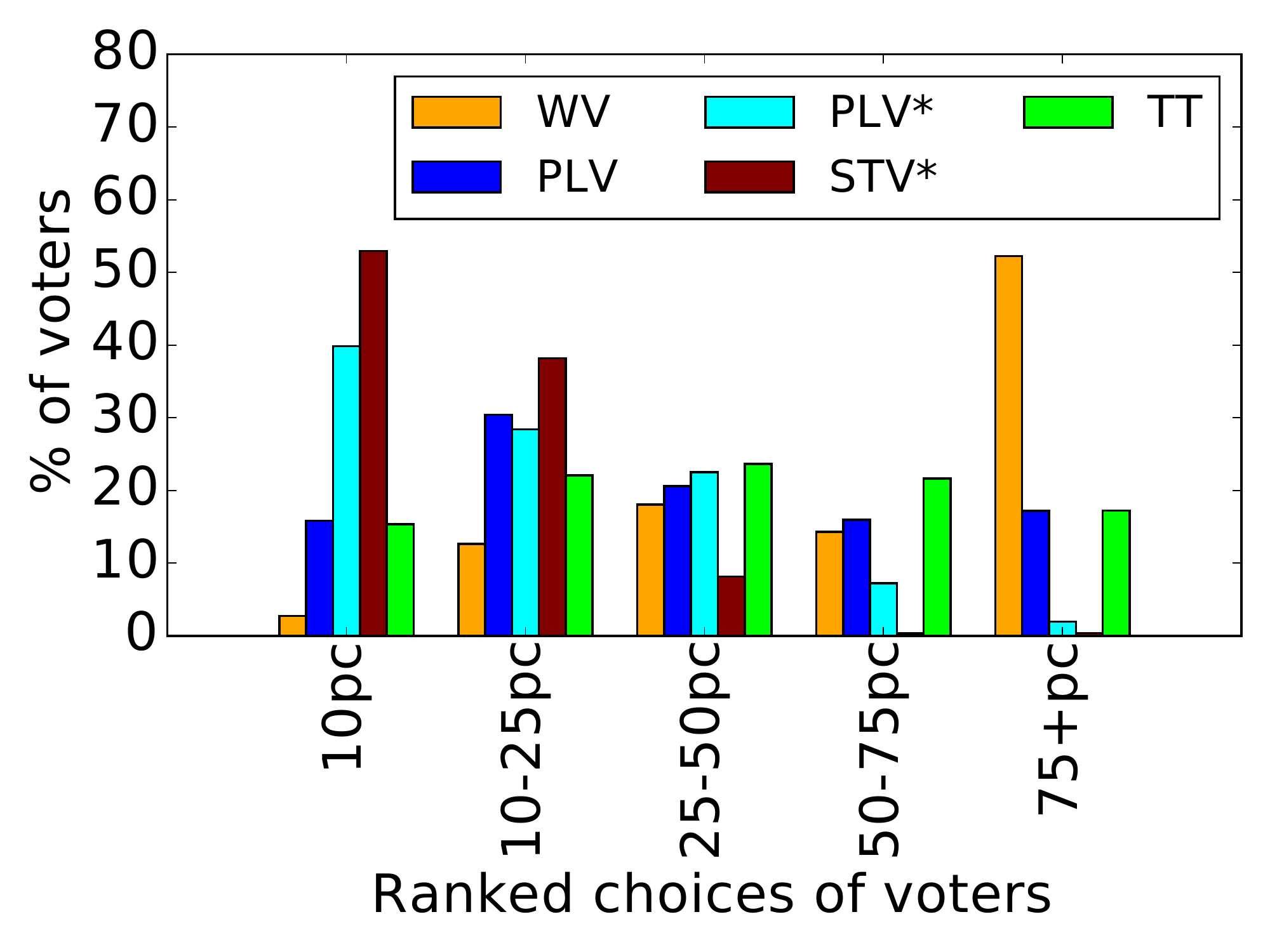}}
}
\vspace*{-4mm}
\caption{{\bf Heatmaps depicting the Jaccard Coefficients between different methods for selecting (a) trending topics in Twitter, and (b) most popular news in Adressa. (c) Average ranked choices and (d) Percentile choices for the trending topics selected by different methods throughout July, 2017.}}
\label{fig:rankedChoice}
\vspace*{-5mm}
\end{figure*}

\vspace{-1mm}
\section{Experimental Evaluation}
\label{sec:evaluation}
In this section, we evaluate the performance of our proposed approaches in selecting items for recommendation. For that, in Twitter, we consider every $15$ minute intervals throughout the month of July, 2017 as election cycles. During any election, from the large set of available hashtags, we select 1,000 candidate hashtags %based on their {\it trendworthiness scores} (which capture the increase in usage of a hashtag 
which experience highest jump in usage during that election cycle (compared to their usage in the previous cycle). While computing the preference rankings of Twitter users, due to Twitter API rate limits, it is not possible to infer ranked choice for everyone. We take $2\%$ random sample from the $15$ million Twitter users in our dataset (resulting in a large sample of $300K$ users)\footnote{The idea of `random voting' is not new. Getting everyone to vote in an election is often impractical or too costly. Dating back to ancient Athenian democracy, philosophers including Aristotle argued for selecting a large random sample of voters and then mandating them to vote during an election~\cite{hansen1991athenian}. 
More recently, Chaum~\cite{chaum2016random} proposed a technique to hold random elections. %As an extension of random election,
Fishkin {\it et al.}~\cite{fishkin2017deliberative} proposed an alternate `Deliberative Polling', where the idea is to select a random sample of voters, give them proper information, ask them to discuss issues and then consider only their votes.}, and gather the ranked choices of all of them (and of no other) over the 1,000 candidates. For the Adressa dataset, we consider every day during February and March, 2017 as  election cycles. We select as candidates top 1,000 stories based on the number of users clicking on them. Then, we compute the preference rankings of all users in our dataset over the candidates. %stories.
After getting the preference rankings of the users, we apply two methods: \\
(i) Consider the preference rankings, and select $K$ items which are the first choice for most users. We denote this method as PLV*, because this is an extension of Plurality Voting described next. \\
(ii) Run STV using the preference rankings and select the winners of the election. Following the convention used for PLV*, in this section, we denote STV as STV* to reflect the fact that the ranked choices of everyone have been considered, not only the active users. \\
Next, we describe the baselines we compare PLV* and STV* against. 

\vspace{-1mm}
\subsection{Baseline approaches}
In addition to the preference rankings, we also gather the votes given by the users participating in  %every candidate tags get during 
an election cycle. Then, using the data, we apply the following approaches to select the winners. \\
\noindent {\bf i. Weighted Voting (WV) : } 
Here, $K$ candidates getting maximum votes win the election regardless of who voted for them and how many times one user voted. Hence, it is vulnerable to manipulation by hyper-active users. \\
\noindent {\bf ii. Plurality Voting (PLV) : } Plurality Voting\footnote{Not to be confused with `Plural Voting', which is a variant of `Weighted Voting'.}, or Single Non-Transferable Vote (SNTV), considers only one vote from a participating user. So, if a particular user voted multiple times, we count only one vote for the candidate she voted the most (with randomly breaking ties). Then, $K$ candidates with maximum %highest number of 
votes win.\\
\noindent {\bf iii. Twitter Trending Topics (TT) : } 
We are not privy to the exact algorithm Twitter employs to select the set of trending topics during an election cycle. Therefore, we consider the algorithm as black-box and compare the hashtags selected by the methods with the hashtags declared as trending by Twitter.

\vspace{-2mm}
\subsection{Quantifying pairwise overlaps}
We first investigate whether these different methods pick very different items or they end up selecting same items during an election. To check that, we gather all the items (i.e., hashtags and news stories) selected by each of the methods, %throughout the one-month period, 
and then compute pairwise overlaps between them. Figure~\ref{fig:rankedChoice}(a) shows the heatmap of Jaccard coefficient between %the tags selected by 
different methods, where Jaccard coefficient between methods $i$ and $j$ is measured as
$\frac{|S_i \cap S_j|}{|S_i \cup S_j|}$, 
where $S_i$ is the set of items selected by method $i$ throughout all election cycles. 

We see from Figure~\ref{fig:rankedChoice}(a) that there is $50\%$ overlap between the trending hashtags selected by PLV* and STV*. TT has around $35\%$ overlap with PLV. There is little overlap between hashtags selected by other methods. Similarly, for Adressa dataset (Figure~\ref{fig:rankedChoice}(b)), we see around $45\%$ overlap between the news stories selected by PLV* and STV*. The rest of the methods do not have much overlap.

\noindent {\bf Takeaway: }  Different methods select mostly different items during election cycles. We only see some common items being selected by our two proposed approaches: PLV* and STV*, possibly because both consider the top preferences of all users. Interestingly, actual Twitter Trending Topics (TT) has the highest overlap with the tags selected by Plurality Voting (PLV). Thus, the Twitter algorithm can be conceptualized as running plurality-based elections. %the . Next, we characterize these differences in finer granularity.

\vspace{-2mm}
\subsection{Comparing ranked choices of users}
Different users have different interests, and thus their ranked choices over different candidate items can vary considerably. 
We now investigate how different election methods capture the choices of these users.
Figure~\ref{fig:rankedChoice}(c) shows, on average, how the hashtags selected by different methods represent different ranked choices of  Twitter users. Figure~\ref{fig:rankedChoice}(d) presents the user choices in different percentile bins. We can observe in both Figure~\ref{fig:rankedChoice}(c) and Figure~\ref{fig:rankedChoice}(d) that the STV* selected tags correspond to top $10$ percentile choices for a majority of users. %In fact, the tags cover top $50$ percentile ranked choices of all the voters; whereas, all other methods select tags that  capture varied preferences. 
While PLV* captures users' top choices to some extent, both PLV and TT appeal very differently to different voters. Finally, WV tends to pick tags which represent top choices of only a few voters and bottom choices for a majority of the voters. \\
{\bf Takeaway: }  STV* consistently selects tags that are the top preferences for all voters; whereas other methods capture both top and bottom preferences, with WV performing the worst. Few actual trends selected by Twitter are least preferred by a lot of voters. We see similar result for the Adressa news data as well.

\begin{figure*}[tb]
\vspace*{-2mm}
\center{
\subfloat[{\bf }]{\includegraphics[width=0.24\textwidth]{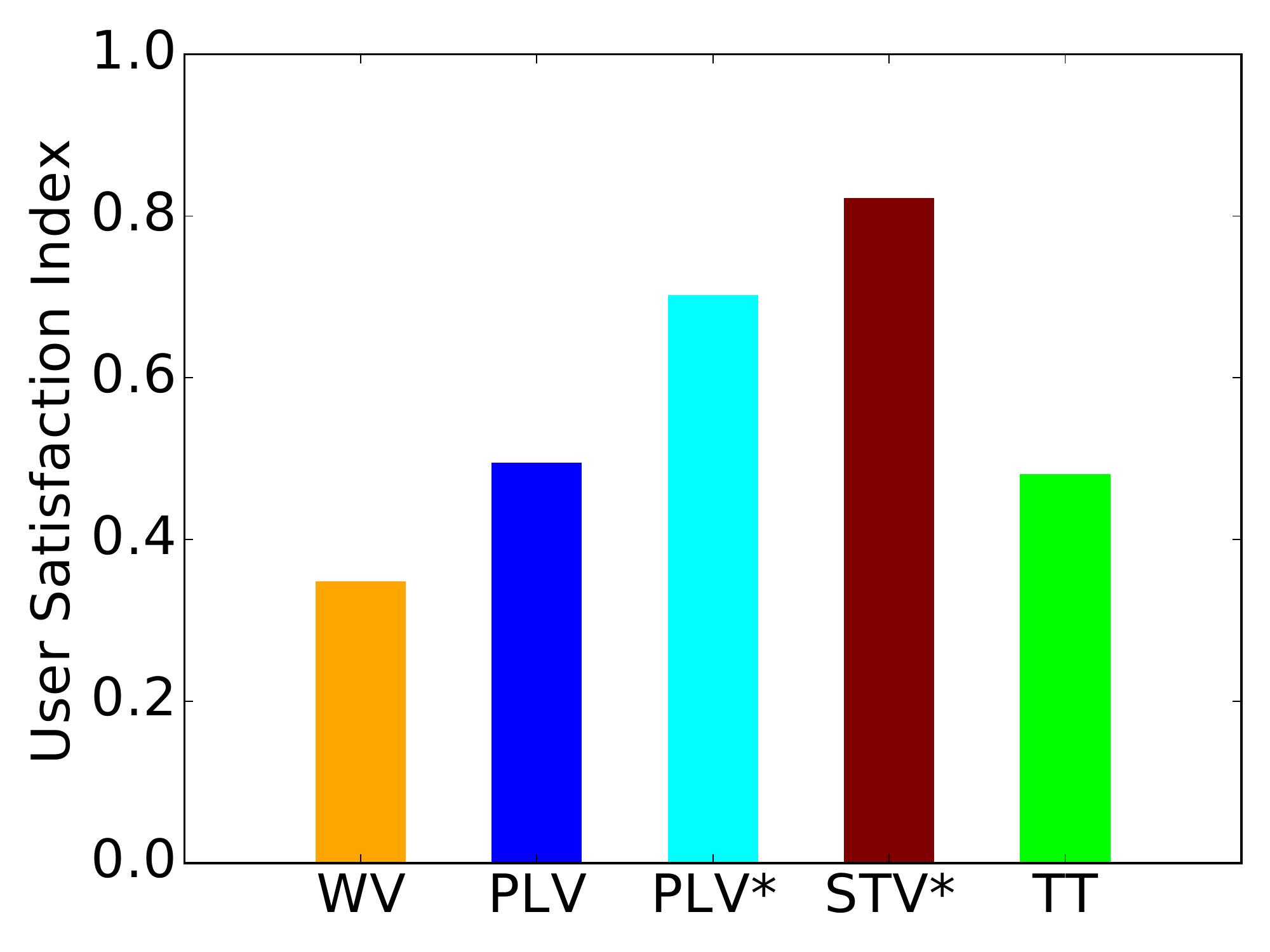}}
\hfil
\subfloat[{\bf }]{\includegraphics[width=0.24\textwidth]{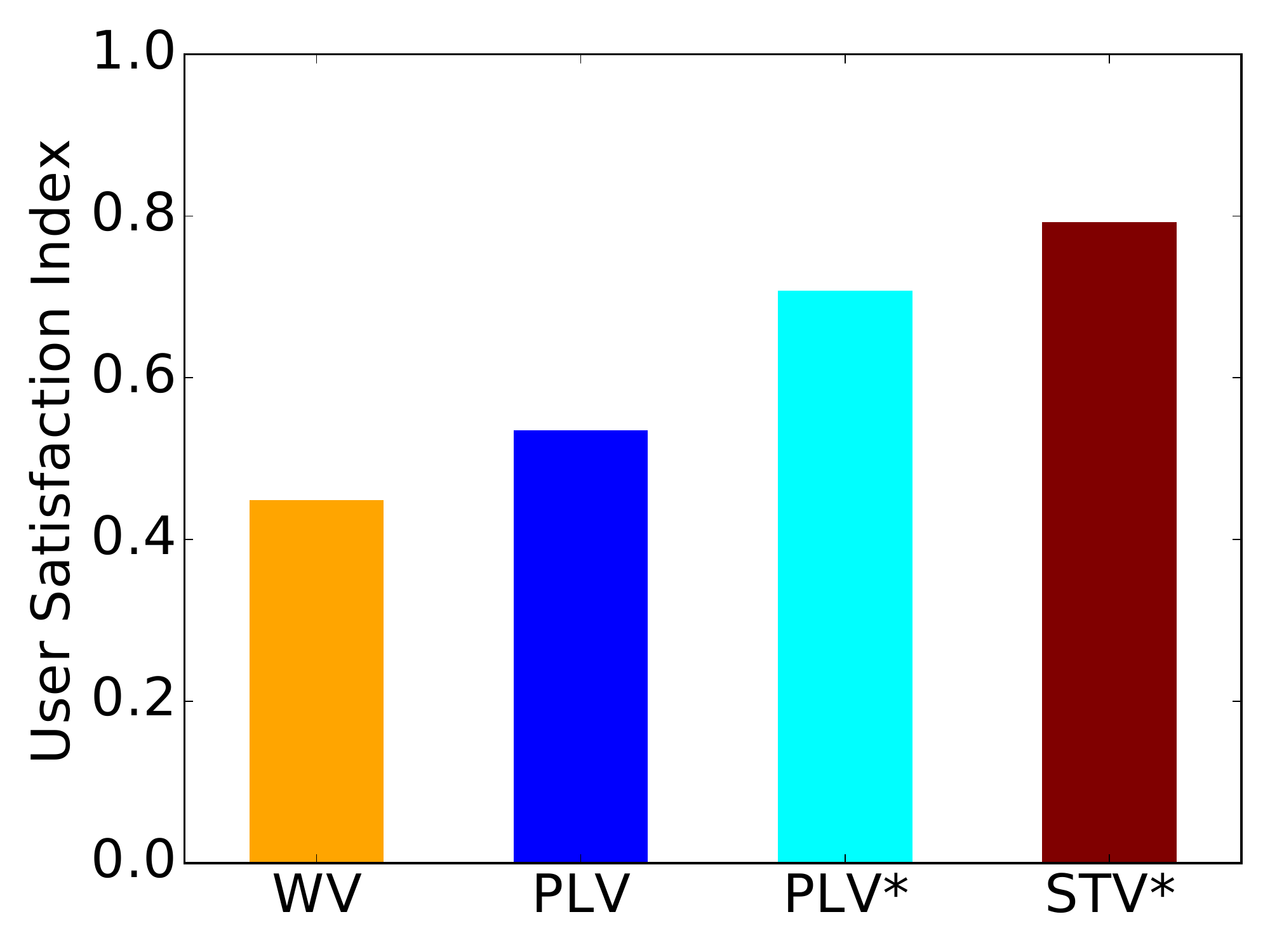}}
\hfil
\subfloat[{\bf }]{\includegraphics[width=0.24\textwidth]{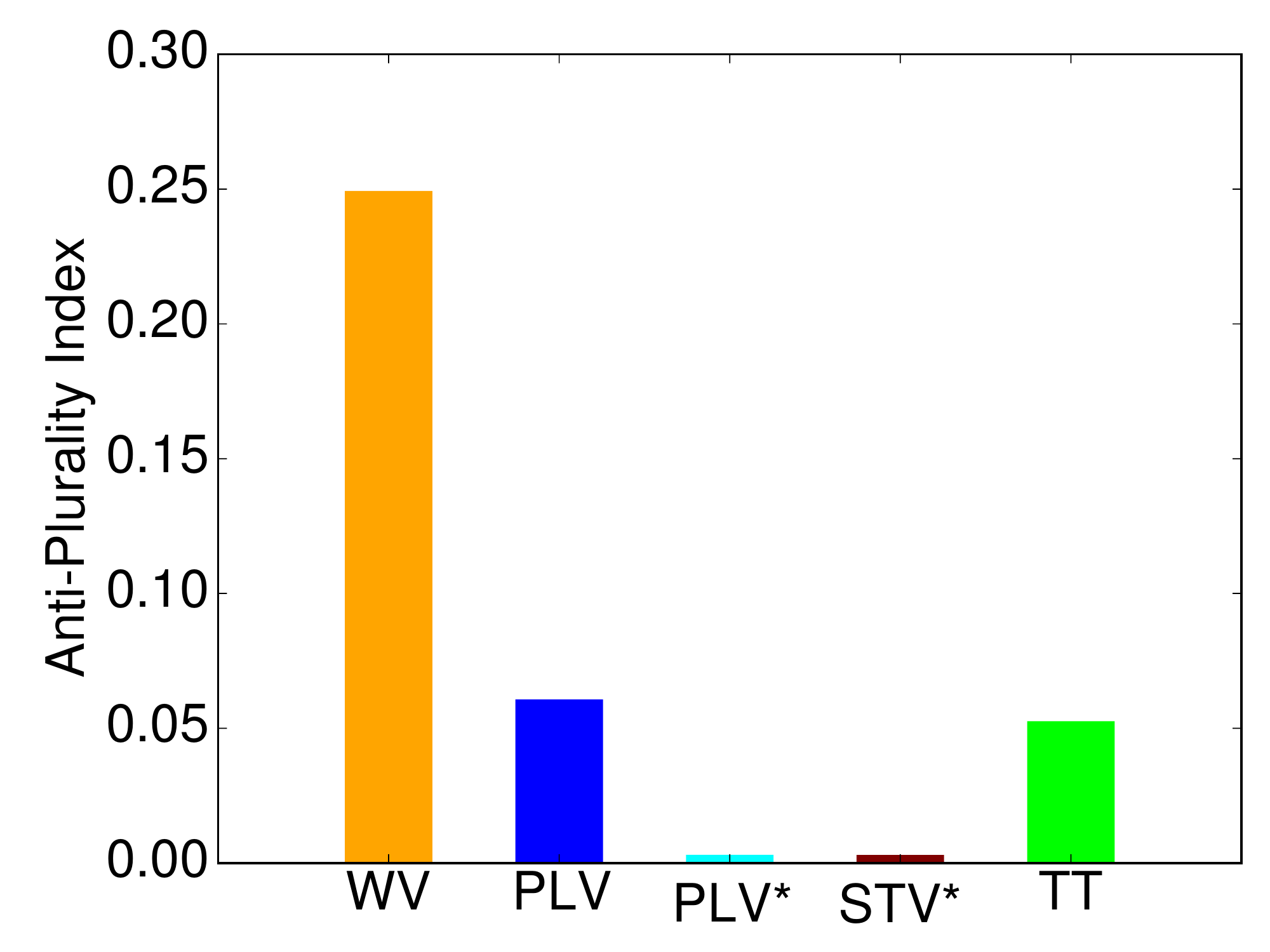}}
\hfil
\subfloat[{\bf }]{\includegraphics[width=0.24\textwidth]{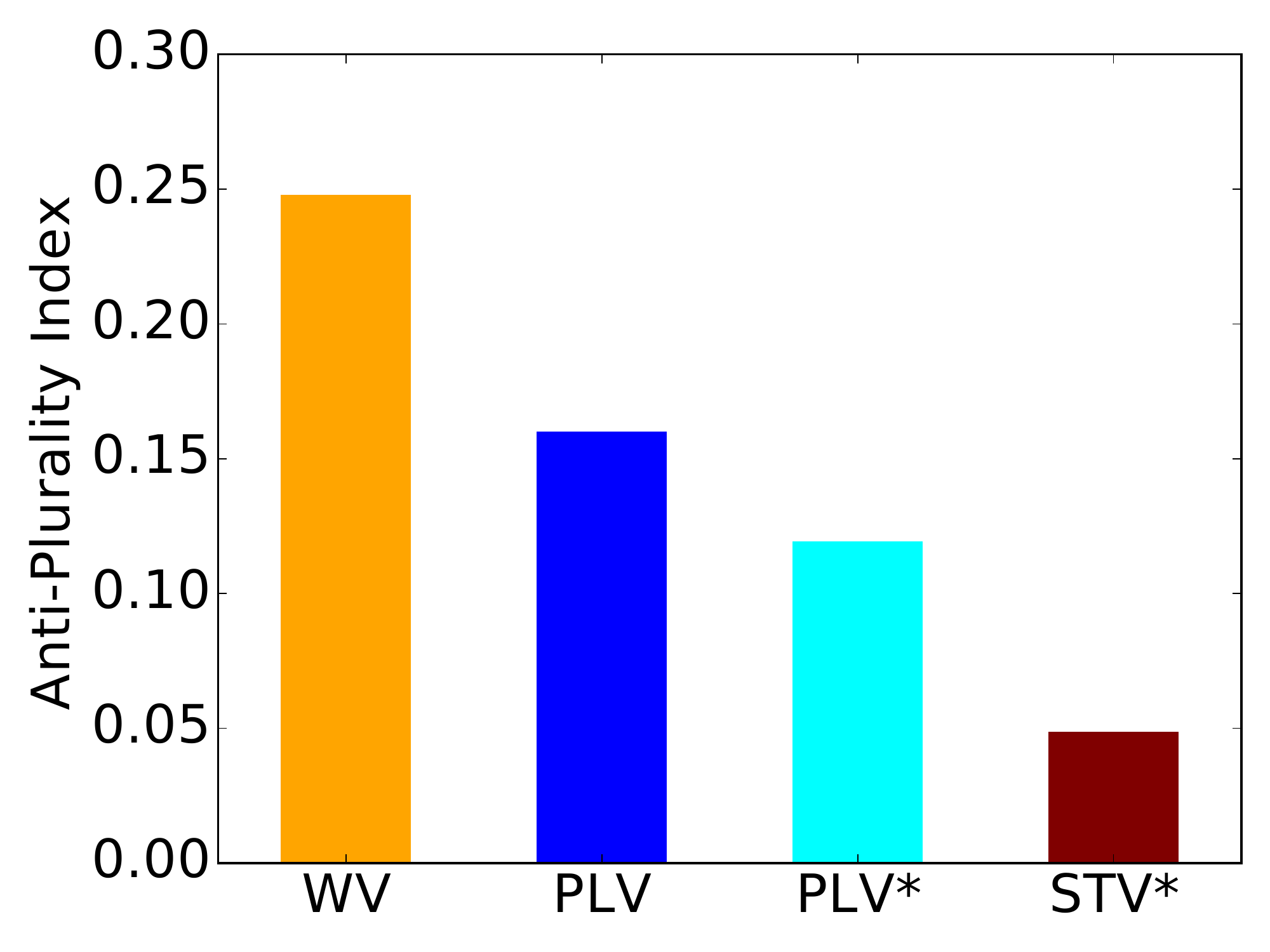}}
}
\vspace*{-4mm}
\caption{{\bf User Satisfaction Index of different methods for computing (a) Twitter Trends, and (b) Most Popular Adressa News. Anti-Plurality Index of different methods for computing (c) Twitter Trends, and (d) Most Popular Adressa News.}}
\label{fig:properties}
\vspace*{-5mm}
\end{figure*}

\vspace{-2mm}
\subsection{Comparing desirable fairness properties}
We now %attempt to succinctly capture how different methods 
compare different methods along the desirable fairness properties identified in Section~\ref{sec:fairness_notion}. 
WV does not satisfy `Equality of Voice' because effectively a voter voting multiple times exerts more power 
than a voter voting once during an election. PLV considers one vote per participating voter; however, it does not consider votes of the silent users. Our proposed PLV* and STV* both guarantee voice equality by giving all users an equal chance to participate in the item selection process.
Regarding the other two properties, we empirically observe to what extent the methods satisfy them.

\subsubsection{\bf User Satisfaction Index}
~\\
The proportional representation criterion requires that if a candidate is preferred by $\frac{1}{K+1}$ fraction of the users, it should be selected, and only STV* theoretically satisfies this criterion. An alternate way to consider representation is from users' perspective. We propose a user  satisfaction criterion which requires that every user should have at least one elected candidate from her top choices. Formally, we consider a user to be satisfied if at least one of its top $10$ choices is selected by a method during an election. Then, {\it User Satisfaction Index} is measured as the fraction of users who are satisfied by a method. Figure~\ref{fig:properties}(a) shows the average User Satisfaction Index for different methods to compute Twitter trends, and we see that both PLV* and STV* are able to satisfy more than $70\%$ of the users; whereas, the other methods cannot satisfy even $50\%$ users. We see similar results for Adressa news dataset as well (Figure~\ref{fig:properties}(b)). 

\subsubsection{\bf Anti-plurality Index}
~\\
The notion of anti-plurality captures whether a method selects items that are disliked by most of the users. We consider a item $i$ to be disliked by a user $u$ if $t$ appears among $v$'s bottom $10$ percentile choices. Then for every such $i$, we compute what percentage of users  dislike $i$ and aggregate this over all the items selected by different methods. Figure~\ref{fig:properties}(c) and Figure~\ref{fig:properties}(d) shows the average Anti-plurality Index for all methods of selecting Twitter Trends and Most Popular Adressa News. We can see in Figure~\ref{fig:properties}(c) that both STV* and PLV* select almost no tags which are disliked by any of the users. On the other hand, WV picks tags which, on average, are disliked by $25\%$ users. For both PLV and TT, the selected tags are disliked by around $5\%$ of all users. Similarly, we can see in Figure~\ref{fig:properties}(d) that STV* has the lowest anti-plurality value (less than $5\%$) while stories selected by WV are disliked by $25\%$ of users.

Specific to Twitter, we observe that there were some extremist tags (e.g., \#exterminate\_syrians, \#IslamIsTheProblem), spammy tags (e.g., \#houstonfollowtrain, \#InternationalEscorts) or politically biased tags (e.g., \#fakeNewsCNN, \#IdiotTrump) which were disliked by more than $90\%$ users, yet got selected by WV or PLV due to the presence of some hyper-active user groups. However, STV* and PLV* did not select any of such hashtags.

\begin{figure*}[tb]
\center{
\subfloat[{\bf }]{\includegraphics[width=0.28\textwidth]{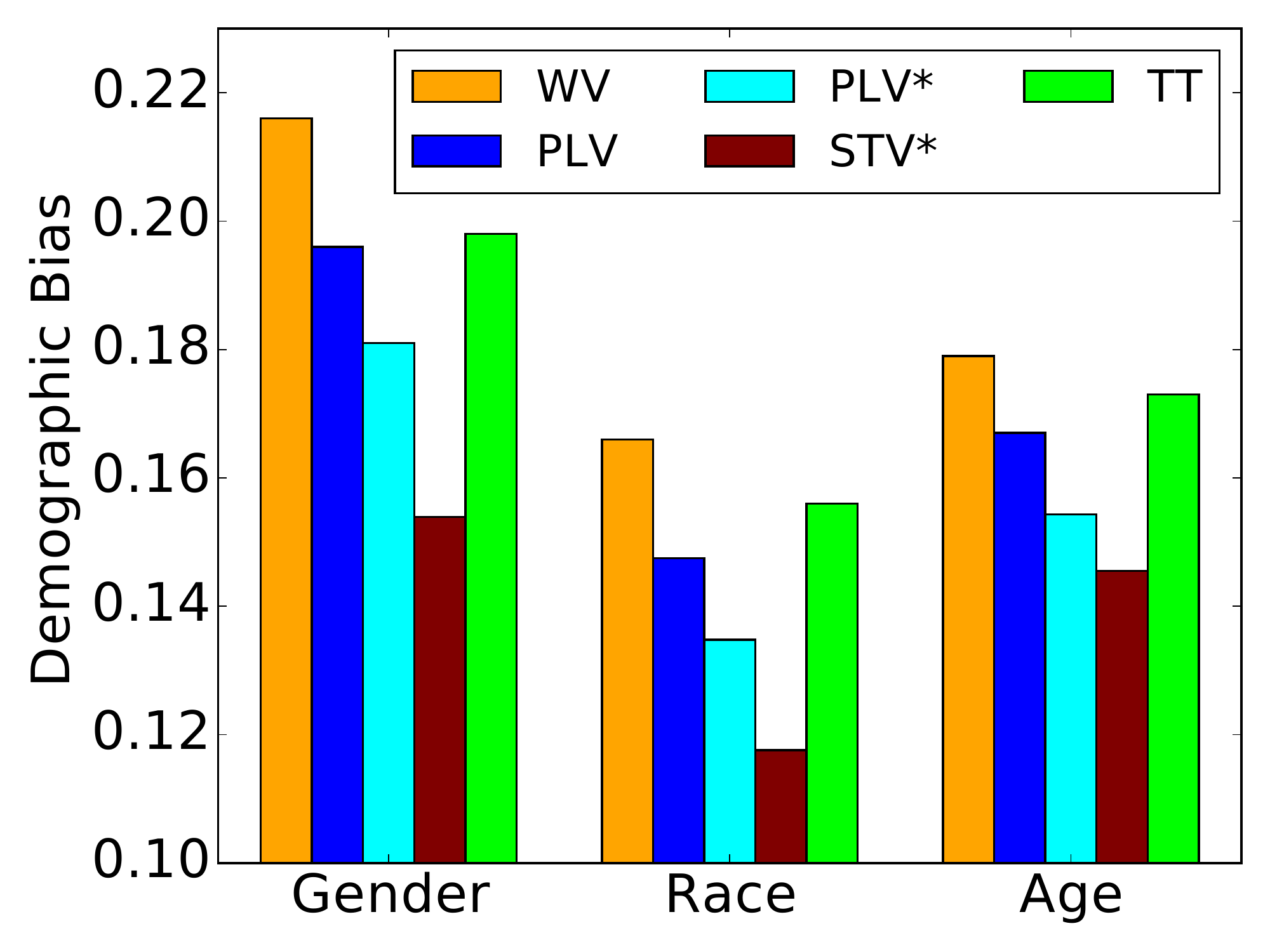}}
\hfil
\subfloat[{\bf }]{\includegraphics[width=0.28\textwidth]{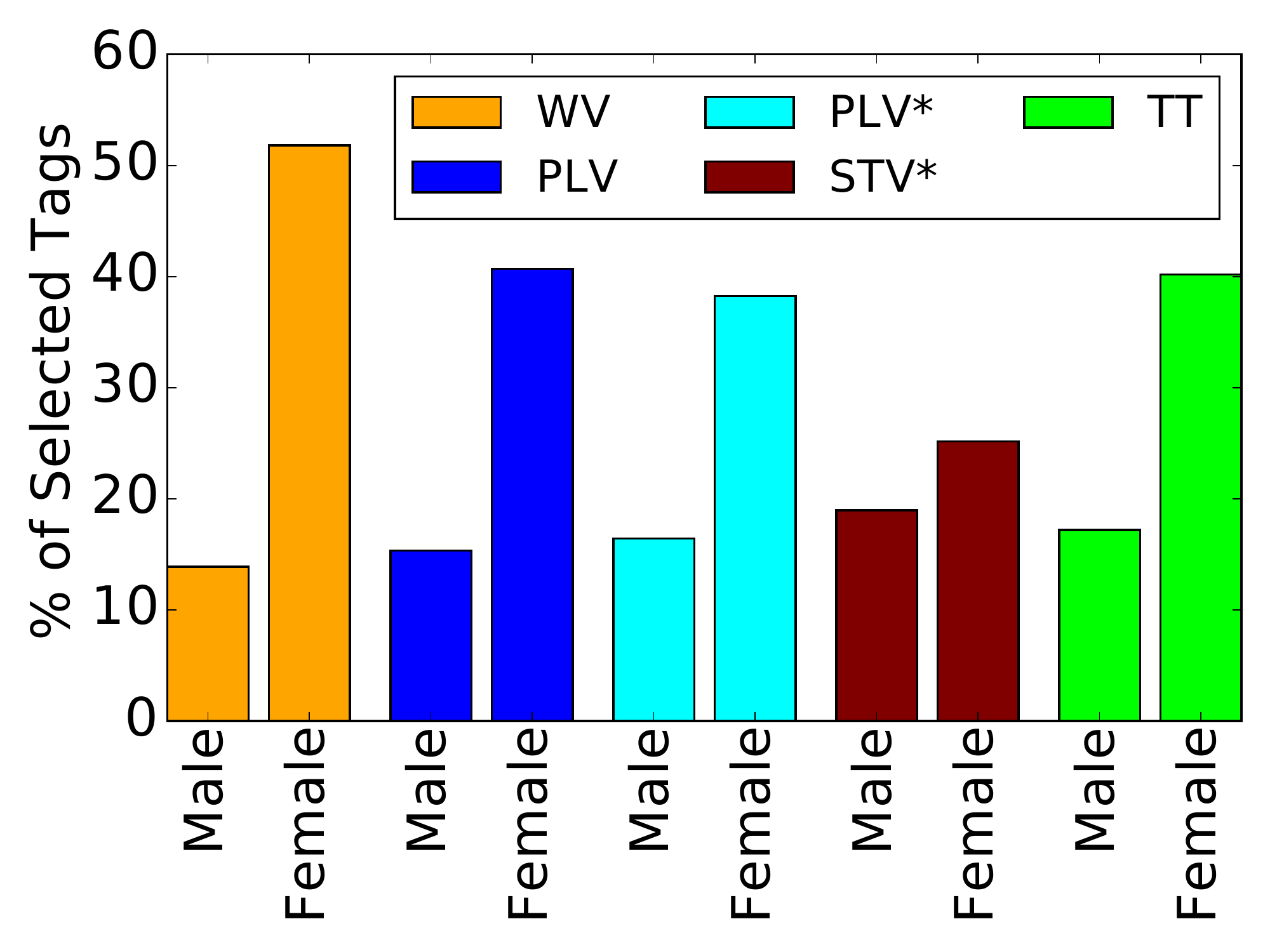}}
\hfil
\subfloat[{\bf }]{\includegraphics[width=0.28\textwidth]{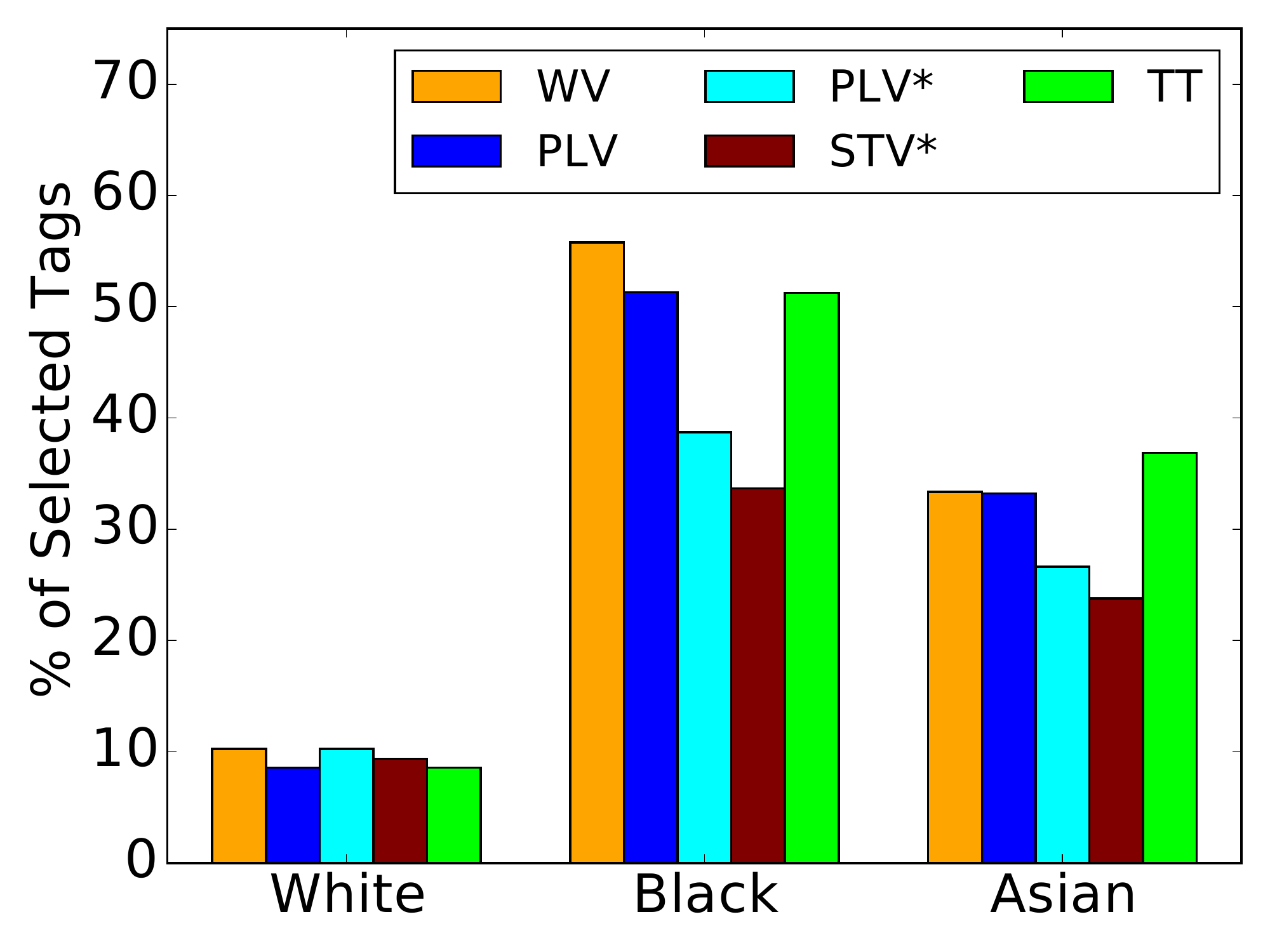}}
}
\vspace*{-2mm}
\caption{{\bf (a) Demographic bias, (b) Gender and (c) Racial under-representation in tags selected by different methods.}}
\label{fig:biasUnderrepresentation}
\vspace*{-3mm}
\end{figure*}

\begin{figure*}[tb]
\center{
\subfloat[{\bf }]{\includegraphics[width=0.28\textwidth]{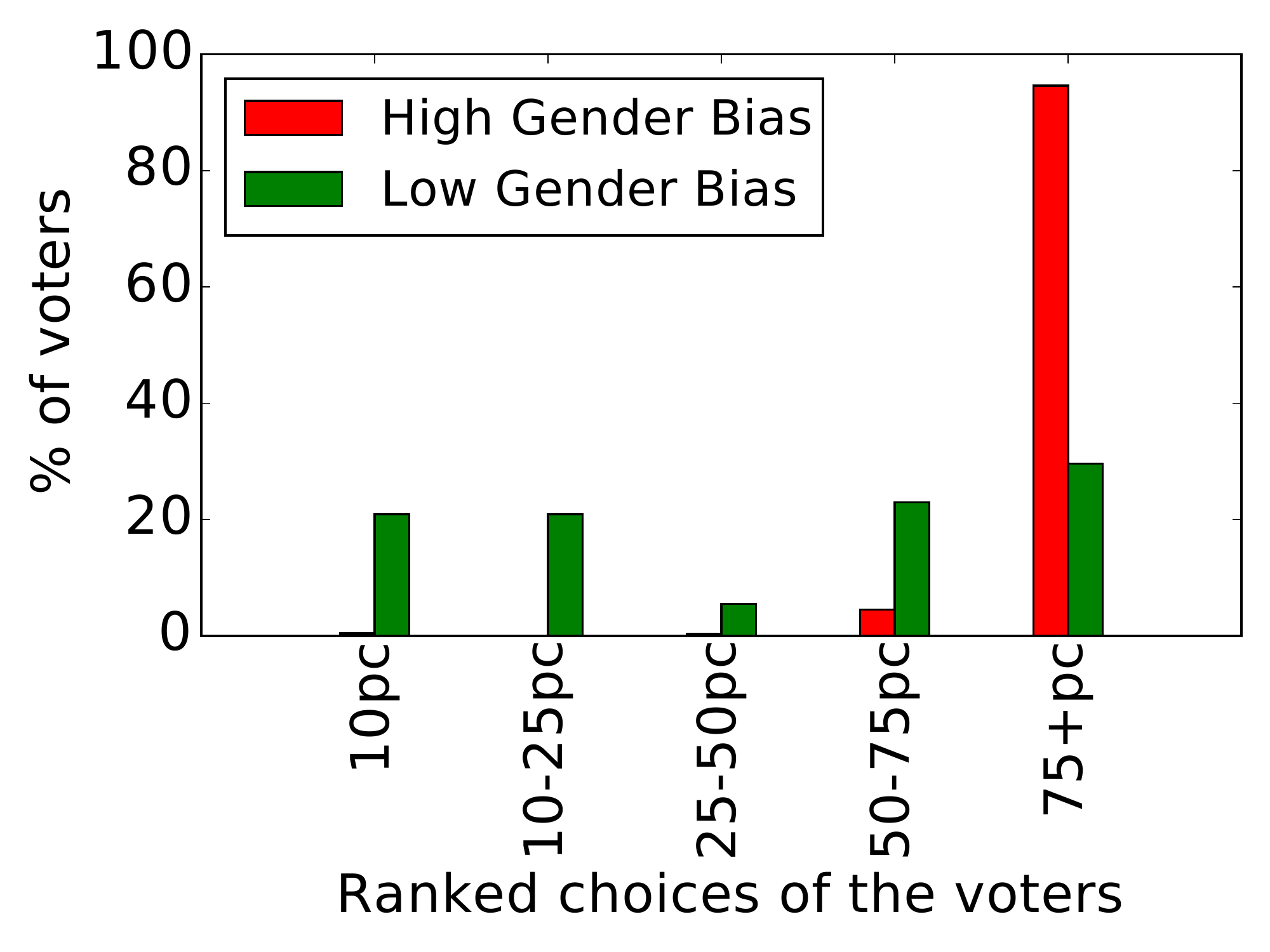}}
\hfil
\subfloat[{\bf }]{\includegraphics[width=0.28\textwidth]{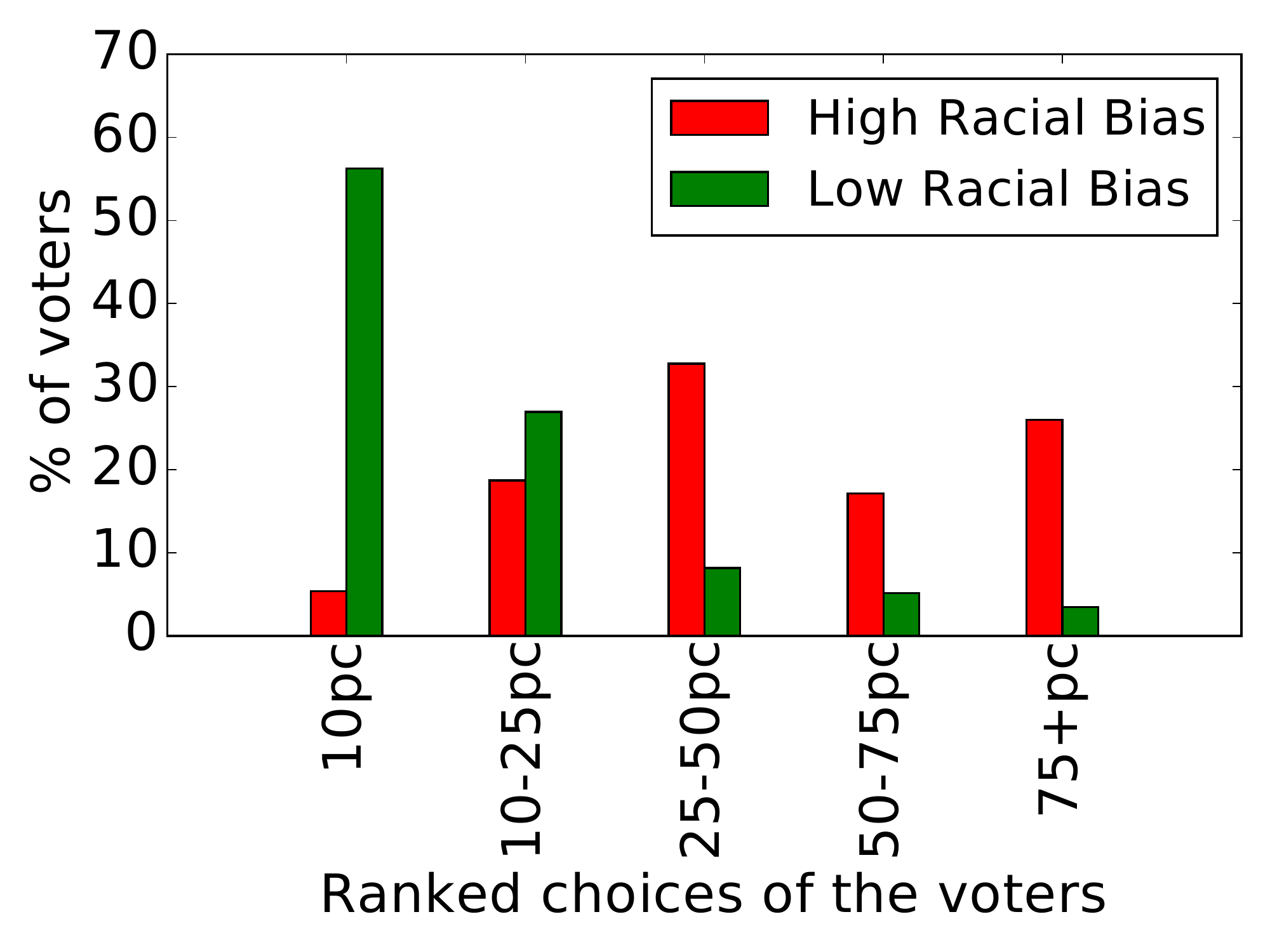}}
\hfil
\subfloat[{\bf }]{\includegraphics[width=0.28\textwidth]{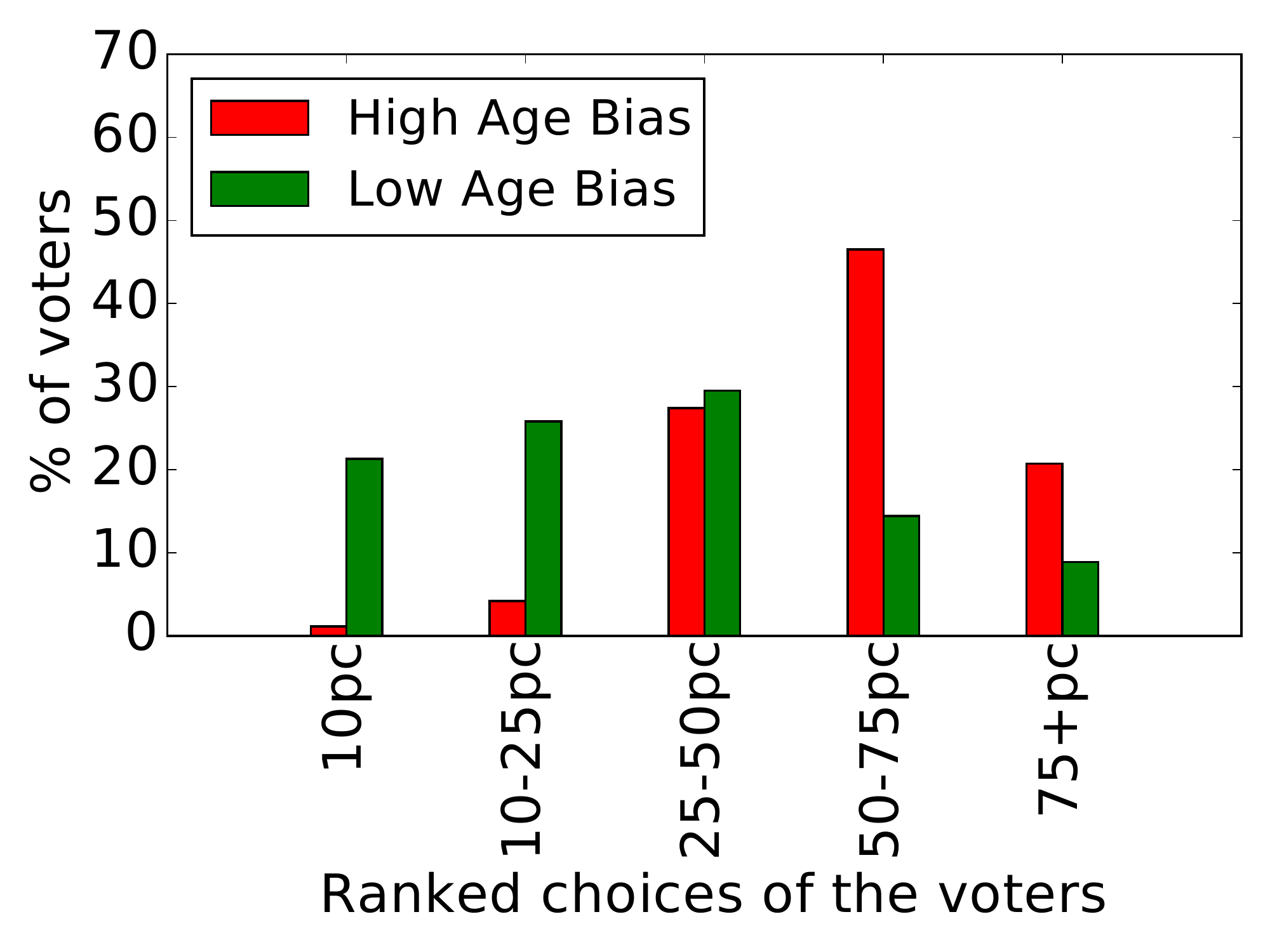}}
}
\vspace*{-2mm}
\caption{{\bf Ranked choices of voters for 100 hashtags most biased and least biased along (a) gender, (b) race and (c) age.}}
\label{fig:biasedTagChoice}
\vspace*{-3mm}
\end{figure*}

\vspace{-2mm}
\subsection{Demographic bias and under-representation in selected topics}
In our earlier work~\cite{chakraborty2017makes}, we found that most of the Twitter trends are promoted by users whose demographics vary significantly from Twitter's overall population. Next, we check whether the voting methods considered in this paper amplify or reduce these demographic biases. We use the demographic information of Twitter users as obtained in~\cite{chakraborty2017makes}. %Following the definition given in~\cite{chakraborty2017makes}, 
Then, demographic bias of tag $i$ is computed as the euclidean distance between the demographics $d_i$ of the people tweeting on $i$ and the reference demographics $d_r$ of the Twitter population in the US~\cite{chakraborty2017makes}:
$
Bias_{i} = || d_i - d_r ||.
$
The higher the score $Bias_{i}$, more biased are the users using the tag $i$.

Figure~\ref{fig:biasUnderrepresentation}(a) shows the average bias across the tags selected by different methods throughout all election cycles.
We see in Figure~\ref{fig:biasUnderrepresentation}(a) that the tags selected by WV are most gender, racially and age biased. On the other hand, STV* selects tags that are least biased. We further observe that considering the preferences of the silent users helps reducing the bias as the average bias of tags selected by PLV* is lower than the average bias of PLV selected tags. 

We next consider the under-representation of different socially salient groups among the users of the tags selected by different methods (where we consider a group $i$ to be under-represented if the fraction of $i$ among the trend users is $< 80\%$ of the fraction of $i$ 
in the overall population~\cite{chakraborty2017makes}).  Figure~\ref{fig:biasUnderrepresentation}(b) shows the under-representation of men and women. In almost all the methods, women are under-represented %among the users of 
for over $40\%$ of the selected tags; whereas, men are under-represented for only around $15\%$ of the tags. However, in the tags selected by STV*, although under-representation of men slightly increases, under-representation of women greatly reduces, having almost equal under-representation of both gender groups. 

Figure~\ref{fig:biasUnderrepresentation}(c) shows the under-representation of different racial groups: Whites, Blacks and Asians. Even though none of the methods achieve similar under-representation of all three racial groups, STV* reduces the under-representation of Blacks and Asians considerably, while keeping the under-representation of Whites similar to other methods. We observe similar trends for age groups where under-representation of Mid-Aged and Adolescents decrease in the tags selected by STV*. The detailed result is omitted for brevity.

\vspace{1mm}
\noindent {\bf How does considering preference rankings reduce \\ demographic bias?} \\
The reduction in demographic bias and under-representation of different social groups among STV* selected tags is surprising because the method has not explicitly taken into account the preference rankings of voters belonging to different demographic groups. We investigate the reason by considering the $100$ most and $100$ least biased tags along all three demographic dimensions -- gender, race and age, and then by checking how they rank in different voters' preference rankings. Figure~\ref{fig:biasedTagChoice} clearly shows that highly biased tags rank low in most of the voter choices. On the other hand, tags with low bias tend to be ranked higher by most of the voters. This interesting observation explains why methods like PLV* or STV* which relies on preference rankings of all the voters %to select the hashtags 
tend to select tags with low bias as compared to other methods like WV or TT which only consider votes by the active users.

\vspace{-1mm}
\section{Related Works}
In this section, we briefly review the related works along two dimensions: top-K item recommendations, and fairness in algorithmic decision making systems.

\vspace{1mm}
\noindent {\bf Top-K recommendations: }
Top-K item recommendations is traditionally associated with personalized recommendation which attempts to find K items a particular user would be mostly interested in~\cite{deshpande2004item}. %The works can be categorized into two broad classes: content-based methods~\cite{lops2011content} and collaborative filtering~\cite{koren2015advances}.
 In content-based recommendations, a user profile is generated based on what she likes or dislikes, and then similar content is identified depending on her past likes~\cite{pazzani2007content}. In collaborative filtering, the preference of a particular user can be inferred based on their similarity to other users~\cite{koren2015advances}. 
However, the recommendation scenario we are considering here is {\it non-personalized}, where the same K items are recommended to everyone. In fact, the problem we are focusing on %in this work 
is how to {\it fairly aggregate personalized preferences of all users} of a website.

\vspace{1mm}
\noindent {\bf Bringing fairness in algorithmic decisions: } Multiple recent works have focused on biases and unfairness in algorithmic decision making~\cite{zafar2017fairness,chakraborty2015can,speicher2018unified}. %The works have categorized algorithmic fairness into two broad categories -- individual and group fairness. Individual fairness ensures that similar individuals are treated similarly and group fairness ensures that there is no disparate treatment on the basis of an individual's belonging to a certain group~\cite{speicher2018unified}. There can be similar kinds of unfairness in personalized and non-personalized recommendations too. 
Yao \textit{et al.}~\cite{yao2017beyond} proposed a few fairness notions for personalized recommendations.  
%{\ few recommendation specific metrics based on proportional representation have also been introduced~\cite{yao2017beyond}. 
Zehlike \textit{et al.}~\cite{zehlike2017fa} introduced fairness in top-k ranking problem through utility based multi-objective formulation.
%, but lacked any application to real world recommender systems. Recently 
Burke~\cite{burke2017multisided} and Chakraborty \textit{et al.}~\cite{chakraborty2017fair} %theoretically 
argued for preserving fairness of consumers (users) as well as suppliers (item producers) in two-sided matching markets. Complementary to  earlier efforts, in this paper, we present the notions of fairness in crowdsourced non-personalized recommendations, and utilize electoral mechanisms to satisfy them in practice.
\vspace{-2mm}
\section{Conclusion and Future Directions}
Recently, there has been a lot of debate and concerns regarding the bias in algorithms operating over big crowd-sourced data. 
In this paper, by conceptualizing crowdsourced recommendation as a multi-winner election, we showed that the bias originates from the unfairness in the electoral process. Then, utilizing long lines of works in social choice theory, we established the fairness properties desired in crowdsourced selections, and identified a particular mechanism STV which satisfies most of these properties. As a result, extensive evaluation over two real-world datasets shows that STV can reduce unfairness and bias in crowdsourced recommendations. Moreover, STV can also resist strategic manipulation by requiring a lot of user support behind potential candidates for recommendation, thereby making it difficult for spammers, bots, or trend hijackers to influence the recommendation process.

There are multiple research directions we want to explore in future. First, our proposed approach can potentially be applied in {\it personalized news recommendation} scenario which combine both user choices and the news trends among the crowds (e.g., Google News~\cite{liu2010personalized}). In such context, at the first level, the candidate stories for recommendation can be selected by standard personalized recommendation algorithms which consider a particular user's interest. Then, an election method like STV can be applied to take into account the crowd choices for electing news stories to recommend to the user. Second, in this work, we conceptualized item (s)election to happen at every fixed intervals; however, there is a streaming component in recommendations like Trending Topics~\cite{mathioudakis2010twittermonitor} (with occasional burstiness in user activities~\cite{kleinberg2003bursty}). Regular election methods are not designed to tackle such scenarios, and we plan to develop mechanisms to handle continuous elections, while simultaneously satisfying the desired fairness properties.

\vspace{1.5mm}
\noindent\textbf{Acknowledgments: }
%\section*{Acknowledgments}
%The authors thank the anonymous reviewers whose suggestions helped to improve the paper. 
This research was supported in part by a European Research Council (ERC) Advanced Grant for the project ``Foundations for Fair Social Computing", funded under the European Union's Horizon 2020 Framework Programme (grant agreement no. 789373). P. Loiseau was supported by the French National Research Agency through the ``Investissements d'avenir" program (ANR-15-IDEX-02) and the Alexander von Humboldt Foundation. A. Chakraborty was a recipient of Google India PhD Fellowship and Prime Minister's Fellowship Scheme for Doctoral Research,
a public-private partnership between Science \& Engineering Research Board (SERB), Department of Science \& Technology, Government of India and Confederation of Indian Industry (CII). 

\balance

%\vspace{-2mm}
{
\small
\bibliographystyle{ACM-Reference-Format}
\bibliography{Main}

%%% -*-BibTeX-*-
%%% Do NOT edit. File created by BibTeX with style
%%% ACM-Reference-Format-Journals [18-Jan-2012].

\begin{thebibliography}{50}

%%% ====================================================================
%%% NOTE TO THE USER: you can override these defaults by providing
%%% customized versions of any of these macros before the \bibliography
%%% command.  Each of them MUST provide its own final punctuation,
%%% except for \shownote{}, \showDOI{}, and \showURL{}.  The latter two
%%% do not use final punctuation, in order to avoid confusing it with
%%% the Web address.
%%%
%%% To suppress output of a particular field, define its macro to expand
%%% to an empty string, or better, \unskip, like this:
%%%
%%% \newcommand{\showDOI}[1]{\unskip}   % LaTeX syntax
%%%
%%% \def \showDOI #1{\unskip}           % plain TeX syntax
%%%
%%% ====================================================================

\ifx \showCODEN    \undefined \def \showCODEN     #1{\unskip}     \fi
\ifx \showDOI      \undefined \def \showDOI       #1{#1}\fi
\ifx \showISBNx    \undefined \def \showISBNx     #1{\unskip}     \fi
\ifx \showISBNxiii \undefined \def \showISBNxiii  #1{\unskip}     \fi
\ifx \showISSN     \undefined \def \showISSN      #1{\unskip}     \fi
\ifx \showLCCN     \undefined \def \showLCCN      #1{\unskip}     \fi
\ifx \shownote     \undefined \def \shownote      #1{#1}          \fi
\ifx \showarticletitle \undefined \def \showarticletitle #1{#1}   \fi
\ifx \showURL      \undefined \def \showURL       {\relax}        \fi
% The following commands are used for tagged output and should be
% invisible to TeX
\providecommand\bibfield[2]{#2}
\providecommand\bibinfo[2]{#2}
\providecommand\natexlab[1]{#1}
\providecommand\showeprint[2][]{arXiv:#2}

\bibitem[\protect\citeauthoryear{Arrow}{Arrow}{1950}]%
        {arrow1950difficulty}
\bibfield{author}{\bibinfo{person}{Kenneth~J Arrow}.}
  \bibinfo{year}{1950}\natexlab{}.
\newblock \showarticletitle{A difficulty in the concept of social welfare}.
\newblock \bibinfo{journal}{\emph{Journal of political economy}}
  \bibinfo{volume}{58}, \bibinfo{number}{4} (\bibinfo{year}{1950}).
\newblock


\bibitem[\protect\citeauthoryear{Baeza-Yates}{Baeza-Yates}{2016}]%
        {baeza2016data}
\bibfield{author}{\bibinfo{person}{Ricardo Baeza-Yates}.}
  \bibinfo{year}{2016}\natexlab{}.
\newblock \showarticletitle{Data and algorithmic bias in the web}. In
  \bibinfo{booktitle}{\emph{WebScience}}.
\newblock


\bibitem[\protect\citeauthoryear{Baker and Potts}{Baker and Potts}{2013}]%
        {baker2013white}
\bibfield{author}{\bibinfo{person}{Paul Baker} {and} \bibinfo{person}{Amanda
  Potts}.} \bibinfo{year}{2013}\natexlab{}.
\newblock \showarticletitle{`Why do white people have thin lips?' Google and
  the perpetuation of stereotypes via auto-complete search forms}.
\newblock \bibinfo{journal}{\emph{Critical Discourse Studies}}
  \bibinfo{volume}{10}, \bibinfo{number}{2} (\bibinfo{year}{2013}).
\newblock


\bibitem[\protect\citeauthoryear{Bartholdi and Orlin}{Bartholdi and
  Orlin}{1991}]%
        {bartholdi1991single}
\bibfield{author}{\bibinfo{person}{John~J Bartholdi} {and}
  \bibinfo{person}{James~B Orlin}.} \bibinfo{year}{1991}\natexlab{}.
\newblock \showarticletitle{Single transferable vote resists strategic voting}.
\newblock \bibinfo{journal}{\emph{Social Choice and Welfare}}
  \bibinfo{volume}{8}, \bibinfo{number}{4} (\bibinfo{year}{1991}).
\newblock


\bibitem[\protect\citeauthoryear{Bhattacharya, Zafar, Ganguly, Ghosh, and
  Gummadi}{Bhattacharya et~al\mbox{.}}{2014}]%
        {bhattacharya2014inferring}
\bibfield{author}{\bibinfo{person}{Parantapa Bhattacharya},
  \bibinfo{person}{Muhammad~Bilal Zafar}, \bibinfo{person}{Niloy Ganguly},
  \bibinfo{person}{Saptarshi Ghosh}, {and} \bibinfo{person}{Krishna~P
  Gummadi}.} \bibinfo{year}{2014}\natexlab{}.
\newblock \showarticletitle{Inferring user interests in the twitter social
  network}. In \bibinfo{booktitle}{\emph{ACM RecSys}}.
\newblock


\bibitem[\protect\citeauthoryear{Burke}{Burke}{2017}]%
        {burke2017multisided}
\bibfield{author}{\bibinfo{person}{Robin Burke}.}
  \bibinfo{year}{2017}\natexlab{}.
\newblock \showarticletitle{Multisided fairness for recommendation}.
\newblock \bibinfo{journal}{\emph{arXiv preprint arXiv:1707.00093}}
  (\bibinfo{year}{2017}).
\newblock


\bibitem[\protect\citeauthoryear{Chakraborty, Ghosh, Ganguly, and
  Gummadi}{Chakraborty et~al\mbox{.}}{2015}]%
        {chakraborty2015can}
\bibfield{author}{\bibinfo{person}{Abhijnan Chakraborty},
  \bibinfo{person}{Saptarshi Ghosh}, \bibinfo{person}{Niloy Ganguly}, {and}
  \bibinfo{person}{Krishna~P Gummadi}.} \bibinfo{year}{2015}\natexlab{}.
\newblock \showarticletitle{Can trending news stories create coverage bias? on
  the impact of high content churn in online news media}. In
  \bibinfo{booktitle}{\emph{Computation and Journalism Symposium}}.
\newblock


\bibitem[\protect\citeauthoryear{Chakraborty, Hannak, Biega, and
  Gummadi}{Chakraborty et~al\mbox{.}}{2017a}]%
        {chakraborty2017fair}
\bibfield{author}{\bibinfo{person}{Abhijnan Chakraborty},
  \bibinfo{person}{Aniko Hannak}, \bibinfo{person}{Asia~J Biega}, {and}
  \bibinfo{person}{Krishna~P Gummadi}.} \bibinfo{year}{2017}\natexlab{a}.
\newblock \showarticletitle{Fair Sharing for Sharing Economy Platforms}.
\newblock  (\bibinfo{year}{2017}).
\newblock


\bibitem[\protect\citeauthoryear{Chakraborty, Messias, Benevenuto, Ghosh,
  Ganguly, and Gummadi}{Chakraborty et~al\mbox{.}}{2017b}]%
        {chakraborty2017makes}
\bibfield{author}{\bibinfo{person}{Abhijnan Chakraborty},
  \bibinfo{person}{Johnnatan Messias}, \bibinfo{person}{Fabricio Benevenuto},
  \bibinfo{person}{Saptarshi Ghosh}, \bibinfo{person}{Niloy Ganguly}, {and}
  \bibinfo{person}{Krishna~P Gummadi}.} \bibinfo{year}{2017}\natexlab{b}.
\newblock \showarticletitle{Who Makes Trends? Understanding Demographic Biases
  in Crowdsourced Recommendations}. In \bibinfo{booktitle}{\emph{AAAI ICWSM}}.
\newblock


\bibitem[\protect\citeauthoryear{Chaum}{Chaum}{2016}]%
        {chaum2016random}
\bibfield{author}{\bibinfo{person}{David Chaum}.}
  \bibinfo{year}{2016}\natexlab{}.
\newblock \bibinfo{title}{Random-sample voting}.
\newblock
\newblock


\bibitem[\protect\citeauthoryear{Deshpande and Karypis}{Deshpande and
  Karypis}{2004}]%
        {deshpande2004item}
\bibfield{author}{\bibinfo{person}{Mukund Deshpande} {and}
  \bibinfo{person}{George Karypis}.} \bibinfo{year}{2004}\natexlab{}.
\newblock \showarticletitle{Item-based top-n recommendation algorithms}.
\newblock \bibinfo{journal}{\emph{ACM TOIS}} \bibinfo{volume}{22},
  \bibinfo{number}{1} (\bibinfo{year}{2004}).
\newblock


\bibitem[\protect\citeauthoryear{Droop}{Droop}{1881}]%
        {droop1881methods}
\bibfield{author}{\bibinfo{person}{Henry~Richmond Droop}.}
  \bibinfo{year}{1881}\natexlab{}.
\newblock \showarticletitle{On methods of electing representatives}.
\newblock \bibinfo{journal}{\emph{Journal of the Statistical Society of
  London}} \bibinfo{volume}{44}, \bibinfo{number}{2} (\bibinfo{year}{1881}).
\newblock


\bibitem[\protect\citeauthoryear{Dummett}{Dummett}{1984}]%
        {Dummett84}
\bibfield{author}{\bibinfo{person}{Michael Dummett}.}
  \bibinfo{year}{1984}\natexlab{}.
\newblock \bibinfo{booktitle}{\emph{Voting procedures}}.
\newblock


\bibitem[\protect\citeauthoryear{Elkind, Faliszewski, Skowron, and
  Slinko}{Elkind et~al\mbox{.}}{2017}]%
        {elkind2017properties}
\bibfield{author}{\bibinfo{person}{Edith Elkind}, \bibinfo{person}{Piotr
  Faliszewski}, \bibinfo{person}{Piotr Skowron}, {and} \bibinfo{person}{Arkadii
  Slinko}.} \bibinfo{year}{2017}\natexlab{}.
\newblock \showarticletitle{Properties of multiwinner voting rules}.
\newblock \bibinfo{journal}{\emph{Social Choice and Welfare}}
  \bibinfo{volume}{48}, \bibinfo{number}{3} (\bibinfo{year}{2017}).
\newblock


\bibitem[\protect\citeauthoryear{Faliszewski, Skowron, Slinko, and
  Talmon}{Faliszewski et~al\mbox{.}}{2017}]%
        {faliszewski2017multiwinner}
\bibfield{author}{\bibinfo{person}{Piotr Faliszewski}, \bibinfo{person}{Piotr
  Skowron}, \bibinfo{person}{Arkadii Slinko}, {and} \bibinfo{person}{Nimrod
  Talmon}.} \bibinfo{year}{2017}\natexlab{}.
\newblock \showarticletitle{Multiwinner voting: A new challenge for social
  choice theory}.
\newblock \bibinfo{journal}{\emph{Trends in Computational Social Choice}}
  (\bibinfo{year}{2017}).
\newblock


\bibitem[\protect\citeauthoryear{Fishkin, Senges, Donahoe, Diamond, and
  Siu}{Fishkin et~al\mbox{.}}{2017}]%
        {fishkin2017deliberative}
\bibfield{author}{\bibinfo{person}{James~S Fishkin}, \bibinfo{person}{Max
  Senges}, \bibinfo{person}{Eileen Donahoe}, \bibinfo{person}{Larry Diamond},
  {and} \bibinfo{person}{Alice Siu}.} \bibinfo{year}{2017}\natexlab{}.
\newblock \showarticletitle{Deliberative polling for multistakeholder internet
  governance: considered judgments on access for the next billion}.
\newblock \bibinfo{journal}{\emph{Information, Comm. \& Society}}
  (\bibinfo{year}{2017}).
\newblock


\bibitem[\protect\citeauthoryear{Ghosh, Sharma, Benevenuto, Ganguly, and
  Gummadi}{Ghosh et~al\mbox{.}}{2012}]%
        {ghosh2012cognos}
\bibfield{author}{\bibinfo{person}{Saptarshi Ghosh}, \bibinfo{person}{Naveen
  Sharma}, \bibinfo{person}{Fabricio Benevenuto}, \bibinfo{person}{Niloy
  Ganguly}, {and} \bibinfo{person}{Krishna Gummadi}.}
  \bibinfo{year}{2012}\natexlab{}.
\newblock \showarticletitle{Cognos: crowdsourcing search for topic experts in
  microblogs}. In \bibinfo{booktitle}{\emph{ACM SIGIR}}.
\newblock


\bibitem[\protect\citeauthoryear{Gulla, Zhang, Liu, {\"O}zg{\"o}bek, and
  Su}{Gulla et~al\mbox{.}}{2017}]%
        {gulla2017adressa}
\bibfield{author}{\bibinfo{person}{Jon~Atle Gulla}, \bibinfo{person}{Lemei
  Zhang}, \bibinfo{person}{Peng Liu}, \bibinfo{person}{{\"O}zlem
  {\"O}zg{\"o}bek}, {and} \bibinfo{person}{Xiaomeng Su}.}
  \bibinfo{year}{2017}\natexlab{}.
\newblock \showarticletitle{The Adressa dataset for news recommendation}. In
  \bibinfo{booktitle}{\emph{ACM WI}}.
\newblock


\bibitem[\protect\citeauthoryear{Hansen}{Hansen}{1991}]%
        {hansen1991athenian}
\bibfield{author}{\bibinfo{person}{Mogens~Herman Hansen}.}
  \bibinfo{year}{1991}\natexlab{}.
\newblock \bibinfo{booktitle}{\emph{The Athenian democracy in the age of
  Demosthenes: structure, principles, and ideology}}.
\newblock


\bibitem[\protect\citeauthoryear{Kendall}{Kendall}{1955}]%
        {kendall1955rank}
\bibfield{author}{\bibinfo{person}{Maurice~G Kendall}.}
  \bibinfo{year}{1955}\natexlab{}.
\newblock \showarticletitle{Rank correlation methods}.
\newblock  (\bibinfo{year}{1955}).
\newblock


\bibitem[\protect\citeauthoryear{Kleinberg}{Kleinberg}{2003}]%
        {kleinberg2003bursty}
\bibfield{author}{\bibinfo{person}{Jon Kleinberg}.}
  \bibinfo{year}{2003}\natexlab{}.
\newblock \showarticletitle{Bursty and hierarchical structure in streams}.
\newblock \bibinfo{journal}{\emph{Data Mining and Knowledge Discovery}}
  \bibinfo{volume}{7}, \bibinfo{number}{4} (\bibinfo{year}{2003}).
\newblock


\bibitem[\protect\citeauthoryear{Koren and Bell}{Koren and Bell}{2015}]%
        {koren2015advances}
\bibfield{author}{\bibinfo{person}{Yehuda Koren} {and} \bibinfo{person}{Robert
  Bell}.} \bibinfo{year}{2015}\natexlab{}.
\newblock \showarticletitle{Advances in collaborative filtering}.
\newblock In \bibinfo{booktitle}{\emph{Recommender systems handbook}}.
  \bibinfo{publisher}{Springer}.
\newblock


\bibitem[\protect\citeauthoryear{Liu, Dolan, and Pedersen}{Liu
  et~al\mbox{.}}{2010}]%
        {liu2010personalized}
\bibfield{author}{\bibinfo{person}{Jiahui Liu}, \bibinfo{person}{Peter Dolan},
  {and} \bibinfo{person}{Elin~R{\o}nby Pedersen}.}
  \bibinfo{year}{2010}\natexlab{}.
\newblock \showarticletitle{Personalized news recommendation based on click
  behavior}. In \bibinfo{booktitle}{\emph{ACM IUI}}.
\newblock


\bibitem[\protect\citeauthoryear{Lops, De~Gemmis, and Semeraro}{Lops
  et~al\mbox{.}}{2011}]%
        {lops2011content}
\bibfield{author}{\bibinfo{person}{Pasquale Lops}, \bibinfo{person}{Marco
  De~Gemmis}, {and} \bibinfo{person}{Giovanni Semeraro}.}
  \bibinfo{year}{2011}\natexlab{}.
\newblock \showarticletitle{Content-based recommender systems: State of the art
  and trends}.
\newblock In \bibinfo{booktitle}{\emph{Rec. Sys. handbook}}.
\newblock


\bibitem[\protect\citeauthoryear{Luo, Zhou, Xia, and Zhu}{Luo
  et~al\mbox{.}}{2014}]%
        {luo2014efficient}
\bibfield{author}{\bibinfo{person}{Xin Luo}, \bibinfo{person}{Mengchu Zhou},
  \bibinfo{person}{Yunni Xia}, {and} \bibinfo{person}{Qingsheng Zhu}.}
  \bibinfo{year}{2014}\natexlab{}.
\newblock \showarticletitle{An efficient non-negative
  matrix-factorization-based approach to collaborative filtering for
  recommender systems}.
\newblock \bibinfo{journal}{\emph{IEEE Transactions on Industrial Informatics}}
  \bibinfo{volume}{10}, \bibinfo{number}{2} (\bibinfo{year}{2014}).
\newblock


\bibitem[\protect\citeauthoryear{Mathioudakis and Koudas}{Mathioudakis and
  Koudas}{2010}]%
        {mathioudakis2010twittermonitor}
\bibfield{author}{\bibinfo{person}{Michael Mathioudakis} {and}
  \bibinfo{person}{Nick Koudas}.} \bibinfo{year}{2010}\natexlab{}.
\newblock \showarticletitle{Twittermonitor: trend detection over the twitter
  stream}. In \bibinfo{booktitle}{\emph{ACM SIGMOD}}. ACM.
\newblock


\bibitem[\protect\citeauthoryear{Mckew}{Mckew}{2018}]%
        {release_the_memo}
\bibfield{author}{\bibinfo{person}{Molly~K. Mckew}.}
  \bibinfo{year}{2018}\natexlab{}.
\newblock \bibinfo{title}{How Twitter Bots and Trump Fans Made \#ReleaseTheMemo
  Go Viral}.
\newblock
  \bibinfo{howpublished}{\url{https://www.politico.com/magazine/story/2018/02/04/trump-twitter-russians-release-the-memo-216935}}.
\newblock


\bibitem[\protect\citeauthoryear{Miller}{Miller}{2003}]%
        {miller2003js}
\bibfield{author}{\bibinfo{person}{J~Joseph Miller}.}
  \bibinfo{year}{2003}\natexlab{}.
\newblock \showarticletitle{JS Mill on plural voting, competence and
  participation}.
\newblock \bibinfo{journal}{\emph{History of political thought}}
  \bibinfo{volume}{24}, \bibinfo{number}{4} (\bibinfo{year}{2003}),
  \bibinfo{pages}{647--667}.
\newblock


\bibitem[\protect\citeauthoryear{Newsroom}{Newsroom}{2016}]%
        {fb_human_removal}
\bibfield{author}{\bibinfo{person}{Facebook Newsroom}.}
  \bibinfo{year}{2016}\natexlab{}.
\newblock \bibinfo{title}{Search FYI: An Update to Trending}.
\newblock
  \bibinfo{howpublished}{\url{https://newsroom.fb.com/news/2016/08/search-fyi-an-update-to-trending/}}.
\newblock


\bibitem[\protect\citeauthoryear{Newsroom}{Newsroom}{2018}]%
        {fb_news_feed}
\bibfield{author}{\bibinfo{person}{Facebook Newsroom}.}
  \bibinfo{year}{2018}\natexlab{}.
\newblock \bibinfo{title}{News Feed FYI}.
\newblock
  \bibinfo{howpublished}{\url{https://newsroom.fb.com/news/category/news-feed-fyi}}.
\newblock


\bibitem[\protect\citeauthoryear{Nunez}{Nunez}{2016}]%
        {fb_bias}
\bibfield{author}{\bibinfo{person}{Michael Nunez}.}
  \bibinfo{year}{2016}\natexlab{}.
\newblock \bibinfo{title}{Former Facebook Workers: We Routinely Suppressed
  Conservative News}.
\newblock
  \bibinfo{howpublished}{https://gizmodo.com/former-facebook-workers-we-routinely-suppressed-conser-1775461006}.
\newblock


\bibitem[\protect\citeauthoryear{Ohlheiser}{Ohlheiser}{2016}]%
        {fb_fake_news}
\bibfield{author}{\bibinfo{person}{Abby Ohlheiser}.}
  \bibinfo{year}{2016}\natexlab{}.
\newblock \bibinfo{title}{Three days after removing human editors, Facebook is
  already trending fake news}.
\newblock
  \bibinfo{howpublished}{{https://www.washingtonpost.com/news/the-intersect/wp/2016/08/29/a-fake-headline-about-megyn-kelly-was-trending-on-facebook}}.
\newblock


\bibitem[\protect\citeauthoryear{Paterek}{Paterek}{2007}]%
        {paterek2007improving}
\bibfield{author}{\bibinfo{person}{Arkadiusz Paterek}.}
  \bibinfo{year}{2007}\natexlab{}.
\newblock \showarticletitle{Improving regularized singular value decomposition
  for collaborative filtering}. In \bibinfo{booktitle}{\emph{ACM SIGKDD}}.
\newblock


\bibitem[\protect\citeauthoryear{Pazzani and Billsus}{Pazzani and
  Billsus}{2007}]%
        {pazzani2007content}
\bibfield{author}{\bibinfo{person}{Michael~J Pazzani} {and}
  \bibinfo{person}{Daniel Billsus}.} \bibinfo{year}{2007}\natexlab{}.
\newblock \showarticletitle{Content-based recommendation systems}.
\newblock In \bibinfo{booktitle}{\emph{The adaptive web}}.
  \bibinfo{publisher}{Springer}.
\newblock


\bibitem[\protect\citeauthoryear{Rendle, Freudenthaler, Gantner, and
  Schmidt-Thieme}{Rendle et~al\mbox{.}}{2009}]%
        {rendle2009bpr}
\bibfield{author}{\bibinfo{person}{Steffen Rendle}, \bibinfo{person}{Christoph
  Freudenthaler}, \bibinfo{person}{Zeno Gantner}, {and} \bibinfo{person}{Lars
  Schmidt-Thieme}.} \bibinfo{year}{2009}\natexlab{}.
\newblock \showarticletitle{BPR: Bayesian personalized ranking from implicit
  feedback}. In \bibinfo{booktitle}{\emph{IUAI}}.
\newblock


\bibitem[\protect\citeauthoryear{Samanta, De, Chakraborty, and Ganguly}{Samanta
  et~al\mbox{.}}{2017}]%
        {samanta2017lmpp}
\bibfield{author}{\bibinfo{person}{Bidisha Samanta}, \bibinfo{person}{Abir De},
  \bibinfo{person}{Abhijnan Chakraborty}, {and} \bibinfo{person}{Niloy
  Ganguly}.} \bibinfo{year}{2017}\natexlab{}.
\newblock \showarticletitle{LMPP: a large margin point process combining
  reinforcement and competition for modeling hashtag popularity}. In
  \bibinfo{booktitle}{\emph{IJCAI}}.
\newblock


\bibitem[\protect\citeauthoryear{Skowron, Faliszewski, and Slinko}{Skowron
  et~al\mbox{.}}{2016}]%
        {skowron2016axiomatic}
\bibfield{author}{\bibinfo{person}{Piotr Skowron}, \bibinfo{person}{Piotr
  Faliszewski}, {and} \bibinfo{person}{Arkadii Slinko}.}
  \bibinfo{year}{2016}\natexlab{}.
\newblock \showarticletitle{Axiomatic characterization of committee scoring
  rules}.
\newblock \bibinfo{journal}{\emph{arXiv preprint arXiv:1604.01529}}
  (\bibinfo{year}{2016}).
\newblock


\bibitem[\protect\citeauthoryear{Solon and Levin}{Solon and Levin}{2016}]%
        {google_bias}
\bibfield{author}{\bibinfo{person}{Olivia Solon} {and} \bibinfo{person}{Sam
  Levin}.} \bibinfo{year}{2016}\natexlab{}.
\newblock \bibinfo{title}{How Google's search algorithm spreads false
  information with a rightwing bias}.
\newblock
  \bibinfo{howpublished}{\url{theguardian.com/technology/2016/dec/16/google-autocomplete-rightwing-bias-algorithm-political-propaganda}}.
\newblock


\bibitem[\protect\citeauthoryear{Speicher, Heidari, Grgic-Hlaca, Gummadi,
  Singla, Weller, and Zafar}{Speicher et~al\mbox{.}}{2018}]%
        {speicher2018unified}
\bibfield{author}{\bibinfo{person}{Till Speicher}, \bibinfo{person}{Hoda
  Heidari}, \bibinfo{person}{Nina Grgic-Hlaca}, \bibinfo{person}{Krishna~P
  Gummadi}, \bibinfo{person}{Adish Singla}, \bibinfo{person}{Adrian Weller},
  {and} \bibinfo{person}{Muhammad~Bilal Zafar}.}
  \bibinfo{year}{2018}\natexlab{}.
\newblock \showarticletitle{A Unified Approach to Quantifying Algorithmic
  Unfairness: Measuring Individual \& Group Unfairness via Inequality Indices}.
  In \bibinfo{booktitle}{\emph{ACM SIGKDD}}.
\newblock


\bibitem[\protect\citeauthoryear{Stafford and Yu}{Stafford and Yu}{2013}]%
        {stafford2013evaluation}
\bibfield{author}{\bibinfo{person}{Grant Stafford} {and}
  \bibinfo{person}{Louis~Lei Yu}.} \bibinfo{year}{2013}\natexlab{}.
\newblock \showarticletitle{An evaluation of the effect of spam on twitter
  trending topics}. In \bibinfo{booktitle}{\emph{IEEE SocialComm}}.
\newblock


\bibitem[\protect\citeauthoryear{Taylor and Zwicker}{Taylor and
  Zwicker}{1992}]%
        {taylor1992characterization}
\bibfield{author}{\bibinfo{person}{Alan Taylor} {and} \bibinfo{person}{William
  Zwicker}.} \bibinfo{year}{1992}\natexlab{}.
\newblock \showarticletitle{A characterization of weighted voting}.
\newblock \bibinfo{journal}{\emph{Proc. of the American mathematical society}}
  \bibinfo{volume}{115}, \bibinfo{number}{4} (\bibinfo{year}{1992}).
\newblock


\bibitem[\protect\citeauthoryear{Twitter}{Twitter}{2010}]%
        {twitter_trends}
\bibfield{author}{\bibinfo{person}{Twitter}.} \bibinfo{year}{2010}\natexlab{}.
\newblock \bibinfo{title}{To Trend or Not to Trend}.
\newblock
  \bibinfo{howpublished}{\url{https://blog.twitter.com/official/en_us/a/2010/to-trend-or-not-to-trend.html}}.
\newblock


\bibitem[\protect\citeauthoryear{Twitter}{Twitter}{2018a}]%
        {twitter_location_filter}
\bibfield{author}{\bibinfo{person}{Twitter}.} \bibinfo{year}{2018}\natexlab{a}.
\newblock \bibinfo{title}{Filtering Tweets by location}.
\newblock
  \bibinfo{howpublished}{\url{https://developer.twitter.com/en/docs/tutorials/filtering-tweets-by-location}}.
\newblock


\bibitem[\protect\citeauthoryear{Twitter}{Twitter}{2018b}]%
        {twitter_rest_api}
\bibfield{author}{\bibinfo{person}{Twitter}.} \bibinfo{year}{2018}\natexlab{b}.
\newblock \bibinfo{title}{Get trends near a location}.
\newblock
  \bibinfo{howpublished}{\url{https://developer.twitter.com/en/docs/trends/trends-for-location/api-reference/get-trends-place}}.
\newblock


\bibitem[\protect\citeauthoryear{VanDam and Tan}{VanDam and Tan}{2016}]%
        {vandam2016detecting}
\bibfield{author}{\bibinfo{person}{Courtland VanDam} {and}
  \bibinfo{person}{Pang-Ning Tan}.} \bibinfo{year}{2016}\natexlab{}.
\newblock \showarticletitle{Detecting hashtag hijacking from twitter}. In
  \bibinfo{booktitle}{\emph{ACM WebScience}}.
\newblock


\bibitem[\protect\citeauthoryear{Woeginger}{Woeginger}{2003}]%
        {woeginger2003note}
\bibfield{author}{\bibinfo{person}{Gerhard~J Woeginger}.}
  \bibinfo{year}{2003}\natexlab{}.
\newblock \showarticletitle{A note on scoring rules that respect majority in
  choice and elimination}.
\newblock \bibinfo{journal}{\emph{Mathematical Social Sciences}}
  \bibinfo{volume}{46}, \bibinfo{number}{3} (\bibinfo{year}{2003}).
\newblock


\bibitem[\protect\citeauthoryear{Yao and Huang}{Yao and Huang}{2017}]%
        {yao2017beyond}
\bibfield{author}{\bibinfo{person}{Sirui Yao} {and} \bibinfo{person}{Bert
  Huang}.} \bibinfo{year}{2017}\natexlab{}.
\newblock \showarticletitle{Beyond parity: Fairness objectives for
  collaborative filtering}. In \bibinfo{booktitle}{\emph{NIPS}}.
\newblock


\bibitem[\protect\citeauthoryear{Zafar, Bhattacharya, Ganguly, Ghosh, and
  Gummadi}{Zafar et~al\mbox{.}}{2016}]%
        {zafar2016wisdom}
\bibfield{author}{\bibinfo{person}{Muhammad~Bilal Zafar},
  \bibinfo{person}{Parantapa Bhattacharya}, \bibinfo{person}{Niloy Ganguly},
  \bibinfo{person}{Saptarshi Ghosh}, {and} \bibinfo{person}{Krishna~P
  Gummadi}.} \bibinfo{year}{2016}\natexlab{}.
\newblock \showarticletitle{On the wisdom of experts vs. crowds: discovering
  trustworthy topical news in microblogs}. In \bibinfo{booktitle}{\emph{ACM
  CSCW}}.
\newblock


\bibitem[\protect\citeauthoryear{Zafar, Valera, Gomez~Rodriguez, and
  Gummadi}{Zafar et~al\mbox{.}}{2017}]%
        {zafar2017fairness}
\bibfield{author}{\bibinfo{person}{Muhammad~Bilal Zafar},
  \bibinfo{person}{Isabel Valera}, \bibinfo{person}{Manuel Gomez~Rodriguez},
  {and} \bibinfo{person}{Krishna~P Gummadi}.} \bibinfo{year}{2017}\natexlab{}.
\newblock \showarticletitle{Fairness beyond disparate treatment \& disparate
  impact: Learning classification without disparate mistreatment}. In
  \bibinfo{booktitle}{\emph{WWW}}.
\newblock


\bibitem[\protect\citeauthoryear{Zehlike, Bonchi, Castillo, Hajian, Megahed,
  and Baeza-Yates}{Zehlike et~al\mbox{.}}{2017}]%
        {zehlike2017fa}
\bibfield{author}{\bibinfo{person}{Meike Zehlike}, \bibinfo{person}{Francesco
  Bonchi}, \bibinfo{person}{Carlos Castillo}, \bibinfo{person}{Sara Hajian},
  \bibinfo{person}{Mohamed Megahed}, {and} \bibinfo{person}{Ricardo
  Baeza-Yates}.} \bibinfo{year}{2017}\natexlab{}.
\newblock \showarticletitle{Fa* ir: A fair top-k ranking algorithm}. In
  \bibinfo{booktitle}{\emph{ACM CIKM}}.
\newblock


\end{thebibliography}
}

\end{document}